\documentclass[12pt]{article}

\usepackage{amsmath}
\usepackage{amssymb}
\usepackage{graphicx}
\usepackage{float}
\usepackage[dvips]{color}
\usepackage{colordvi}
\usepackage{url}
\usepackage{wrapfig}

\allowdisplaybreaks
\def\Tr{{\rm Tr}}

\begin{document}

\thispagestyle{empty}
\addtocounter{page}{-1}
{}
\vskip-5cm
\begin{flushright}
KEK-TH-1558\\
CAS-KITPC/ITP-326\\
\end{flushright}
\vspace*{0.1cm} \centerline{\Large \bf New States of Gauge Theories on a Circle}
\vspace*{0.1cm}
\vspace*{1 cm} 
\centerline{\bf 
Takehiro~Azuma$^1$, Takeshi~Morita$^2$ and Shingo~Takeuchi$^3$}
\vspace*{0.5cm}
\centerline{\rm $^1$\it Institute for Fundamental Sciences, Setsunan University}
\centerline{\it 17-8 Ikeda Nakamachi, Osaka, 572-8508,  \rm JAPAN}
\vspace*{0.3cm}
\centerline{\rm $^2$\it KEK Theory Center}
\centerline{\it High Energy Accelerator Research Organization (KEK)}
\centerline{\it Ibaraki 305-0801, \rm JAPAN}
\vspace*{0.3cm}
\centerline{\rm $^3$\it State Key Laboratory of Theoretical Physics}
\centerline{\it  Institute of Theoretical
Physics, Chinese Academy of Sciences}
\centerline{\it  Beijing 100190, \rm CHINA}
\vspace*{0.3cm}
\centerline{\tt email: azuma(at)mpg.setsunan.ac.jp, tmorita(at)post.kek.jp}
\centerline{\tt shingo(at)itp.ac.cn}

\vspace*{0.4cm}
\centerline{\bf Abstract}
\vspace*{0.3cm} 

\enlargethispage{1000pt}

We study a one-dimensional large-$N$ $U(N)$ gauge theory on a circle as a toy model of higher dimensional Yang-Mills theories at finite temperature.
To investigate the profile of the thermodynamical potential in this model, we evaluate a stochastic time evolution of several states, and find that an unstable confinement phase at high temperature does not decay to a stable deconfinement phase directly.
Before it reaches the deconfinement phase, it develops to several intermediate states.
These states are characterised by the expectation values of the Polyakov loop operators, which wind the temporal circle different times.
We reveal that these intermediate states are the saddle point solutions of the theory, and similar solutions exist in a wide class of $SU(N)$ and $U(N)$ gauge theories on $S^1$ including QCD and pure Yang-Mills theories in various dimensions.
We also consider  a Kaluza-Klein gravity, which is the gravity dual of the one-dimensional gauge theory on a spatial $S^1$, and show that these solutions may be related to multi black holes localised on the $S^1$.
Then we present a connection between the stochastic time evolution of the gauge theory and the dynamical decay process of a black string though the Gregory-Laflamme instability.

\newpage

\section{Introduction}

Understanding of the thermodynamics of gauge theories is a significant subject in theoretical physics  \cite{Gross:1980br}
.
In the case of the pure Yang-Mills theory, due to the developments of the lattice gauge theory \cite{Creutz:1984mg}
and the holography in string theory \cite{Maldacena:1997re, Witten:1998zw, Gross:1998gk, Aharony:1998qu,Mandal:2011ws}, its several properties have been confirmed.
According to these studies, the theory is confined at low temperature and deconfined at high temperature, and a phase transition happens between these two phases.

However we have not understood the natures sufficiently. 
Although the confinement and deconfinement phases are stable at low and high temperatures respectively, it is unclear whether all unstable and meta-stable states, like domain walls and plasma balls \cite{Aharony:2005bm}, have been derived.
Besides, if such states exist, the profile of the potential, which connects these states, should be investigated to understand the dynamics such as the decay of these states to the stable phase.

In this article, in order to investigate these problems, we employ a ``stochastic time evolution", which is a kind of Monte Carlo calculation introduced in Sec.~\ref{subsec-stochastic}.
We will apply this method to a finite temperature one-dimensional gauge theory at large $N$, which we will explain in the next paragraph, as a toy model for higher dimensional Yang-Mills theories.
Then we evaluate the stochastic time evolution of unstable states, e.g. a deconfinement configuration at low temperature and a confinement configuration at high temperature.
Interestingly we find that the unstable confinement phase at high temperature does not decay to a stable deconfinement phase directly.
It develops to several intermediate states before it reaches the deconfinement phase.
This evolution reflects certain characteristic aspects of the thermodynamical potential of this model.
We will reveal that these intermediate states are related to the saddle point solutions of the potential, and similar solutions exist in a wide class of $SU(N)$ and $U(N)$  gauge theories on a $S^1$ circle including QCD and pure Yang-Mills theories in various dimensions.
Thus these saddle point solutions may characterise the thermodynamical properties of these gauge theories too.

Now we introduce the above one-dimensional $U(N)$ gauge theory, which we mainly study in this article. The action is defined by
\begin{align} 
S=\int_0^\beta \! dt \Tr \left( 
\sum_{I=1}^D \frac{1}{2} \left( D_t Y^I \right)^2
-\sum_{I,J}\frac{g^2}{4} [Y^I,Y^J]^2
\right).
\label{BFSS}
\end{align} 
Here $Y^I$ is adjoint scalar $(I=1,\cdots, D)$ and $D_t =\partial_t - i[A_t, \cdot ]$.
$\beta$ and $A_t$ denote the inverse temperature $1/T$ and the gauge field on the Euclidean time circle respectively.
This model is a Kaluza-Klein reduction of a $(D+1)$-dimensional pure Yang-Mills theory on $S^1_{\beta}\times T^D$ for small $T^D$ \cite{Luscher:1982ma, Mandal:2011hb}.

We study this model at large $N$ \cite{'tHooft:1973jz}. 
We have three motivations to consider the large-$N$ limit. 
Firstly a large-$N$ phase transition occurs in this model, which is akin to the confinement/deconfinement phase transition in higher dimensional pure Yang-Mills theories  \cite{Mandal:2011hb,Aharony:2004ig,Aharony:2005ew,Mandal:2009vz}.
Hence this model would be regarded as a toy model for the Yang-Mills theories.
Secondly we can consider the dual gravity at large $N$ \cite{Aharony:2004ig,Itzhaki:1998dd,Harmark:2004ws}.
Lastly, since the large-$N$ limit is a kind of semi-classical limit, the role of the saddle point solutions would be emphasised.

As we mentioned, we investigate this model by using the stochastic time evolution, and find the new saddle point solutions.
In addition, we will also analyse them  by using a $1/D$ expansion \cite{Mandal:2009vz, Hotta:1998en} and reveal several relevant properties of these solutions including the critical temperatures, below which the solutions cannot exist.
Besides we will compare the observables of these solutions from the Monte Carlo calculation and the $1/D$ expansion, and confirm good agreements.

We will also study an application to the gauge/gravity correspondence  \cite{Aharony:2004ig, Itzhaki:1998dd, Harmark:2004ws}.
If we regard the temporal circle of the model (\ref{BFSS}) as a spatial $S^1$, this model at $D=9$ can be obtained as an effective theory of $N$ D1 branes winding this spatial $S^1$ at high temperature.
Then we can construct the dual geometry by using these D1 branes \cite{Aharony:2004ig}\footnote{Correctly, we will take a T-duality along the $S^1$ and consider D0 brane geometries instead of the D1 branes as we argue in Sec.~\ref{sec-gravity}.}.
Since the geometry involves the spatial $S^1$, the gravity becomes a  Kaluza-Klein theory.
We will argue that the saddle point solutions in the gauge theory would correspond to multi black branes localised on the $S^1$.
Indeed the existence of the dual phases in the gauge theories corresponding to these multi black branes in the Kaluza-Klein theories has been predicted in Ref.~\cite{Harmark:2004ws}, and we will provide evidence for this conjecture.

In addition, it has been shown in Refs.~\cite{Choptuik:2003qd, Lehner:2010pn, Lehner:2011wc} that the multi black holes on $S^1$ (connected by thin black strings) appear during a decay process of a black string due to the Gregory-Laflamme instability \cite{Gregory:1994bj}.
We will argue that this phenomenon may be related to the stochastic time evolution of the unstable confinement phase of the  gauge theory (\ref{BFSS})\footnote{Indeed one motivation of this article is to understand the black string decay through the gauge theory, which is related to a ongoing project \cite{BMMW}. }.

Such a connection to the gravity indicates that the stochastic time evolution must capture relevant properties of the large-$N$ gauge theories on $S^1$. 
Since such gauge theories on $S^1$ are studied in various contexts: finite temperature gauge theories \cite{Gross:1980br}, higher dimensional extensions of the standard model in the phenomenology of particle physics \cite{Manton:1979kb, Fairlie:1979at, Hosotani:1983xw} and string theories, we believe that our results will illuminate some non-perturbative aspects of these theories, especially in the dynamical problems.
Concrete applications will be discussed in Sec.~\ref{sec-conclusion}.

This article is organised as follows.
Although we find the saddle point solutions through the stochastic time evolution in the model (\ref{BFSS}), we first derive these solutions in four dimensional Yang-Mills theory and QCD through a weak coupling analysis in section \ref{sec-QCD}.
This argument will provide a basic idea of the saddle point solutions.
In section \ref{sec-1d}, we study the model (\ref{BFSS}) by using the $1/D$ expansion, and reveal the natures of the saddle points in non-perturbative regime.
In section \ref{sec-cascade}, we investigate the stochastic time evolution of the model (\ref{BFSS}) and show that the saddle points indeed appear as the intermediate states between the unstable confinement phase and the stable deconfinement phase at high temperature.
In section \ref{sec-gravity}, we argue the role of these solutions in the gauge/gravity correspondence.
Section \ref{sec-conclusion} contains the conclusions and discussions.
In appendix \ref{app-multi}, we show some details of the analysis of the model (\ref{BFSS}).
In appendix \ref{app-MC}, we explain the details of our Monte Carlo calculation.

\section{Saddle Point Solutions in Gauge Theories}
\label{sec-QCD}

In this section,  we study  saddle point solutions in $SU(N)$ and $U(N)$ gauge theories at finite temperature.
The application to the gauge theories on a spatial $S^1$ is straightforward\footnote{We have to take care of the periodicities of the fermions along the circle. It is fixed anti-periodic along the temporal circle, whereas we have to choose it either periodic or anti-periodic  along the spatial circle.}. 

To investigate the gauge theories, 
the static diagonal gauge \cite{Gross:1980br}
\begin{align} 
(A_{t})_{ij} = \alpha_i \delta_{ij}/\beta,  \quad (i,j=1,\cdots,N)
\label{diagonal}
\end{align} 
is useful.
Here $A_t$ is the temporal component of the gauge field, and $\alpha_i$ does not depend on the Euclidean time $t$ and has a periodicity $2\pi$.
If the theory is $SU(N)$, they satisfy a constraint $\sum_{i=1}^N \alpha_i = 0$ (mod $ 2\pi$), although it is irrelevant at large $N$, which we will study in later sections.

The phases of the gauge theories are characterised by the expectation value of the Polyakov loop operator 
\begin{align} 
P \equiv \frac{1}{N}  \Tr {\rm P} \exp\left(i \int_{0}^{\beta} A_{t} dt  \right)= \frac{1}{N}\sum_{k=1}^{N} e^{i \alpha_k}
.
\label{Poly-loop}
\end{align} 
In adjoint theories, the phase is confined if $\langle |P| \rangle=0$
 and  is deconfined if $\langle | P| \rangle \neq  0$ \cite{Gross:1980br}. 

In the static diagonal gauge (\ref{diagonal}), the configuration of the eigenvalue $\{\alpha_k\}$ determines the value of the Polyakov loop operator (\ref{Poly-loop}).
Since $\{\alpha_k\}$ can be regarded as the positions of $N$ particles on $S^1$, this $N$ particle problem governs the phase structures of the gauge theories.
Then we can interpret the mechanism of the confinement and deconfinement as follows.
If repulsive forces between the eigenvalue $\{\alpha_k\}$ are dominant, the eigenvalues tend to spread on the $S^1$ and the stable configuration would be
\begin{align} 
\alpha_k=2\pi k/N+c \quad ({\rm mod}~ 2\pi),
\label{confinement}
\end{align} 
and their permutations.
Here $c$ is a $k$-independent constant, which is discretised as $2\pi n/N$ ($n \in Z_N$) in the $SU(N)$ case. 
Then $|P|=0$ is satisfied and the confinement phase is realised.

Oppositely, if attractive forces are dominant, the eigenvalues tend
 to clump and the configuration
\begin{align} 
\alpha_k=c  \quad ({\rm mod}~ 2\pi)
\label{deconfinement}
\end{align} 
would be stable.
Sometimes quantum effects disturb this configuration and the eigenvalue distribution is smeared around $c$ as we will see in Sec.~\ref{sec-1d}.
In this configuration, $|P| \neq 0 $ is satisfied, and the deconfinement phase is realised.
Therefore the phases in the gauge theories may be related to the forces for the eigenvalue $\{\alpha_k\}$.

Then we intuitively notice that, if the attractive forces are dominant at certain temperature, the configurations with multiple mobs of the eigenvalues  may be possible as a solution of the theory, if the attractive forces between the mobs are balanced.
We call such a solution ``multi-peak solution'' and will find these solutions in several models\footnote{The multi peaks of the eigenvalues of $A_t$ are characterised by the expectation values of the generalised Polyakov loop (\ref{un}) too.}.

\subsection{Four Dimensional Pure Yang-Mills Theory}
\label{subsec-YM}

We consider four dimensional pure Yang-Mills theory at high temperature.
 If we assume $\{ \alpha_i \}$ does not depend on the spatial positions, we obtain the one-loop effective potential for $\{ \alpha_i \}$, which would be valid at high temperature $T \gg \Lambda_{{\rm QCD}}$  \cite{Gross:1980br},
\begin{align} 
V_{{\rm eff}}=&\frac{\pi^2 T^4}{24} \sum_{j,k}^N \left[1-\left( 
\left(\frac{\alpha_j}{\pi}-\frac{\alpha_k}{\pi} \right)_{{\rm mod}~ 2}-1
\right)^2 \right]^2. 
\label{qcd-pot}
\end{align} 
This potential implies that the attractive forces between the eigenvalues work, and indeed it is minimised by the configuration (\ref{deconfinement}).
Hence the system is deconfined at high temperature as is well known.

\begin{figure}
\begin{center}
\begin{tabular}{cc}
\begin{minipage}{0.5\hsize}
\begin{center}
\includegraphics[scale=.3]{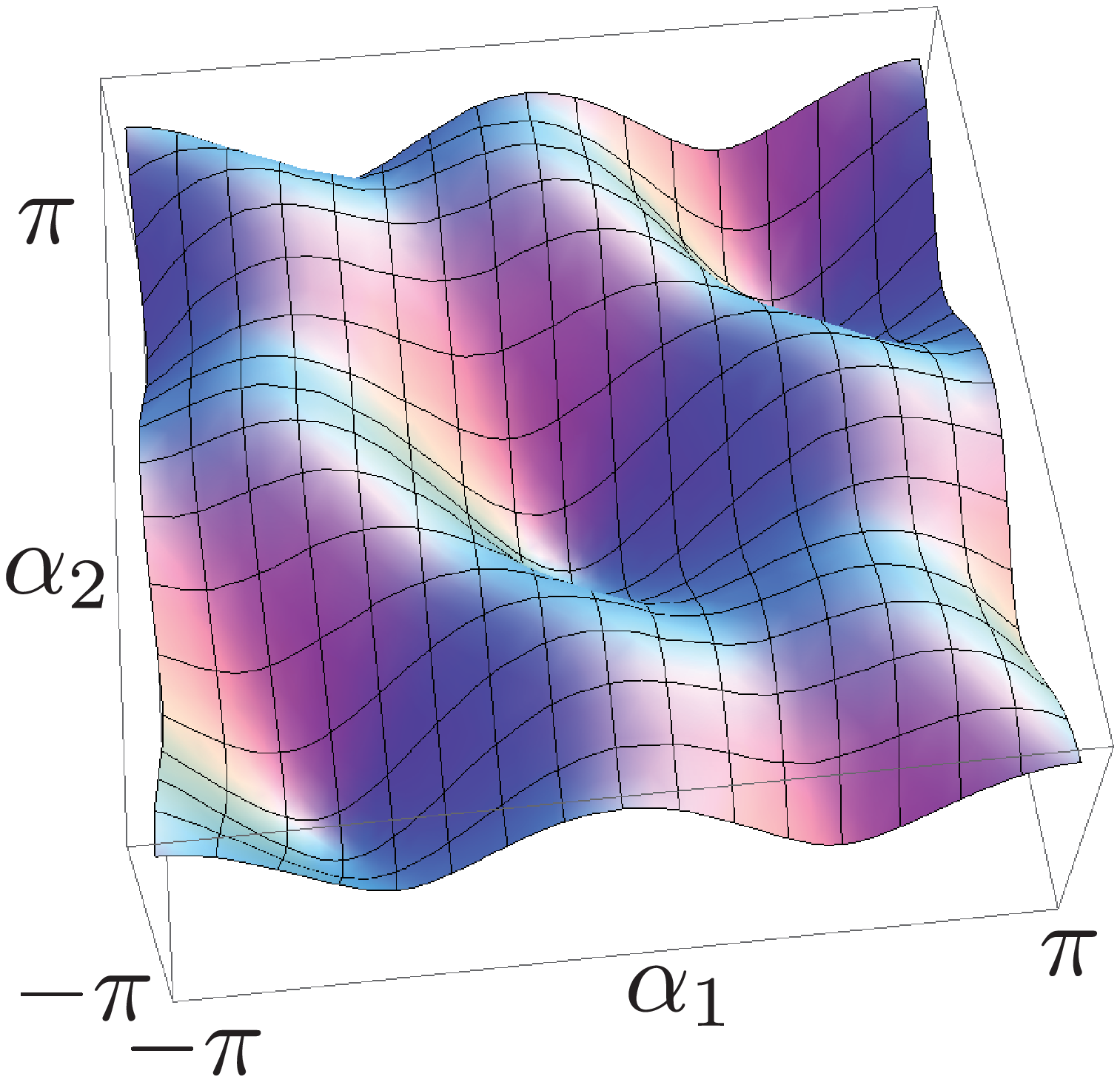}
\end{center}
\end{minipage}
\begin{minipage}{0.5\hsize}
\begin{center}
\includegraphics[scale=.3]{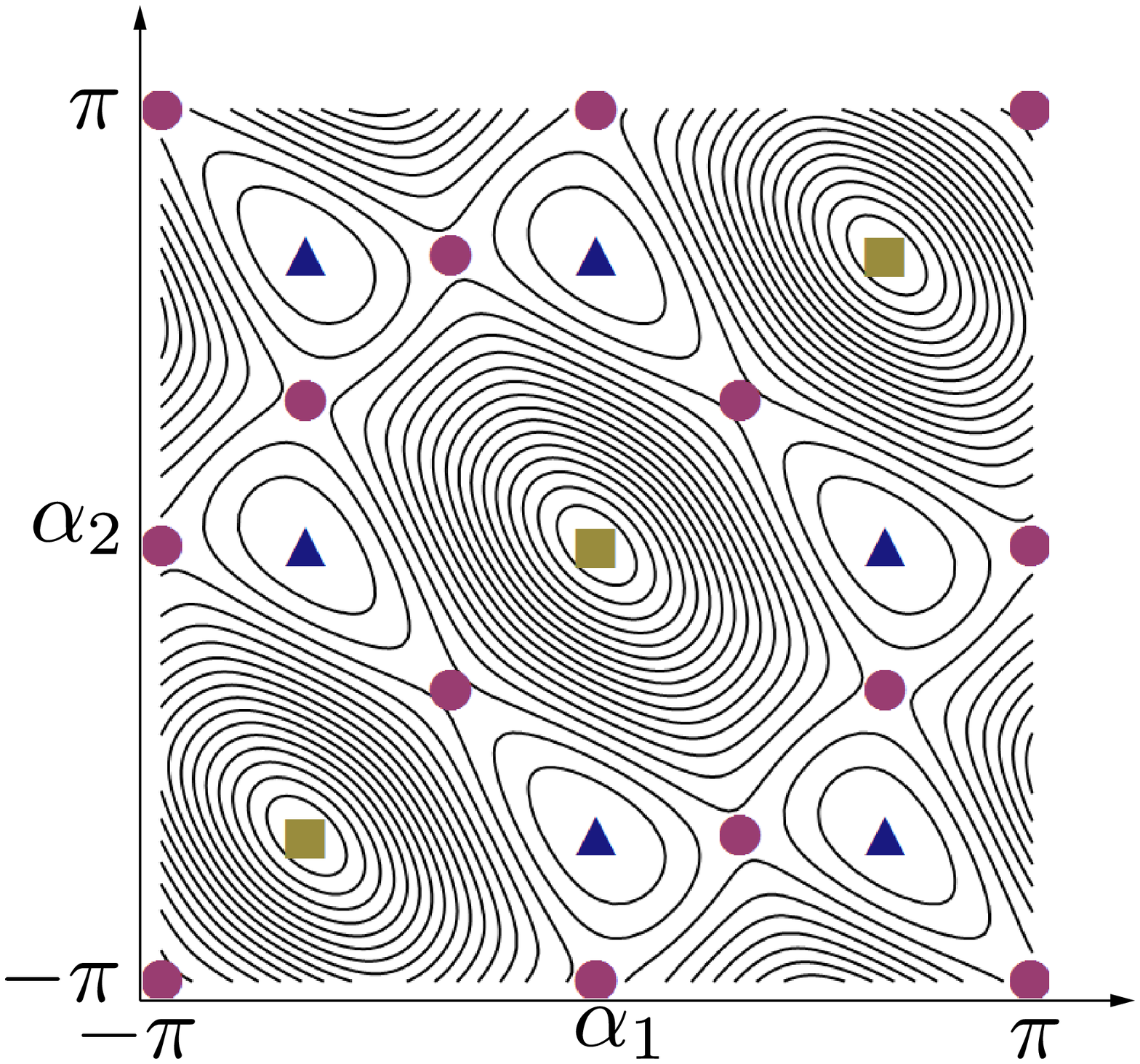}
\end{center}
\end{minipage}
\end{tabular}
\caption{The effective potential (\ref{qcd-pot}) for $SU(3)$.
We fix $\alpha_3= - \alpha_1-\alpha_2$ (mod $2\pi$) and plot (\ref{qcd-pot}) as a function of $\alpha_1$ and $\alpha_2$.
The left is the three dimensional plot and the right is its contour line.
In the right, $\blacksquare$, $\blacktriangle$ and $\bullet$ denote the minimum, maximum and saddle point of the potential, respectively.
}
\label{fig-SU3}
\end{center}
\end{figure}

We plot the potential (\ref{qcd-pot}) for $SU(3)$ in Fig.~\ref{fig-SU3}. 
Then we notice that this potential has several maxima and saddle points in addition to the vacuum solution (\ref{deconfinement}).
The maxima are given by the confinement configuration (\ref{confinement}).
The saddle points are given by $\{\alpha_k\}_{k=1,2,3}=\{\pi+2\pi n/3,\pi+2\pi n/3,2\pi n/3\}$ ($n \in Z_3$) and their permutations.
At these saddle points, the eigenvalues are antipodal on the $S^1$, and the attractive forces are balanced.

\begin{figure}
\begin{center}
\includegraphics[scale=1]{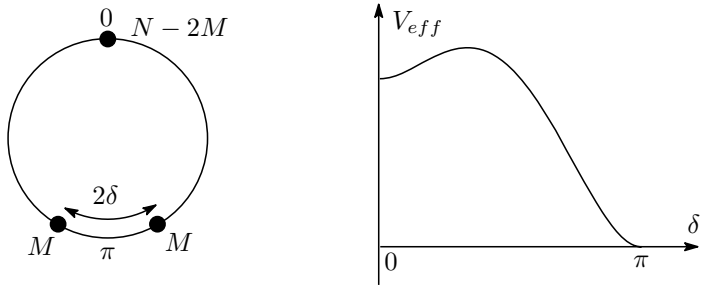}
\caption{The two-peak solution of the potential (\ref{qcd-pot}).
We put $N-2M$ eigenvalues at $0$ and  $M$ eigenvalues at $ \pi \pm \delta$.
The right plot is the potential for $\delta$ ($0 \le \delta \le \pi$).
This plot displays that the two-peak solution ($\delta=0$) is meta-stable against this perturbation and is a saddle point of the potential (\ref{qcd-pot}). 
}
\label{fig-QCD}
\end{center}
\end{figure}

Such multi-peak saddle points generally exist in $SU(N)$ and $U(N)$ Yang-Mills theories if $N\ge 3$.
We show it by considering a two-peak solution as an example.
We arrange $N-2M$ eigenvalues  at $0$ and $2M$ eigenvalues  at $\pi$, and then this configuration becomes a solution of the action (\ref{qcd-pot}), since the attractive forces are balanced.
Although this configuration is obviously unstable,
it is meta-stable against certain $Z_2$ symmetric perturbations.
For instance, if we separate the $2M$ eigenvalues at $\pi$ to $M$ eigenvalues at $\pi \pm  \delta $, then they tend to go back to $\pi$ for a small $\delta$ as shown in Fig.~\ref{fig-QCD}. 
Thus this two-peak configuration is a saddle point of the effective potential (\ref{qcd-pot}).
Similarly we can find various saddle point solutions by adjusting the positions of the eigenvalues.

Particularly, if $N$ is a multiple of an integer $m$, we can construct a $Z_m$ symmetric saddle point solution,
\begin{align} 
\alpha_i= 2\pi k/m +c, \quad  (k-1)N/m < i \le kN/m, \quad (k=1,2,\cdots,m).
\label{Zm}
\end{align} 
We call such a $Z_m$ symmetric solution ``$Z_m$ solution'' in this article\footnote{Similar $Z_m$ symmetric solutions may appear as stable states in gauge theories on  $S^1$ coupled to adjoint fermions with the periodic boundary condition \cite{Unsal:2010qh}. }.
Note that the uniform configuration (\ref{confinement}) is a special case of this solution with $m=N$ but it is unstable against any perturbation and is a maximum rather than a saddle point of the potential (\ref{qcd-pot}).

\subsection{Four Dimensional QCD}
\label{subsec-QCD}

We consider four dimensional QCD by adding $N_f$ fundamental quarks to the  pure Yang-Mills theory which we have argued in the previous subsection.
In this case, the situation is slightly modified.
The $N_f$ fundamental quarks induce the one-loop potential for $\{ \alpha_k \}$ \cite{Gross:1980br},
\begin{align} 
-&\frac{N_f \pi^2 T^4}{12} \sum_{j=1}^N \left[1-\left( 
\left(\frac{\alpha_j}{\pi}+1 \right)_{{\rm mod}~ 2}-1
\right)^2 \right]^2 ,
\end{align} 
in addition to the effective potential (\ref{qcd-pot}).
This potential attracts the eigenvalues at $\alpha_k= 0$ (mod $2\pi$), and breaks the shift symmetry of $\alpha_k$. 
Even in this case, we can adjust the positions of the eigenvalues to obtain multi-peak saddle point solutions.

\subsection{Large-$N$ Gauge Theories}
\label{subsec-general}
%
%
We generalise the saddle point solutions in the previous subsections to large-$N$ gauge theories.
At large $N$, the difference between $SU(N)$ and $U(N)$ is irrelevant in our study and we ignore it.

\begin{figure}
\begin{center}
\includegraphics[scale=1.3]{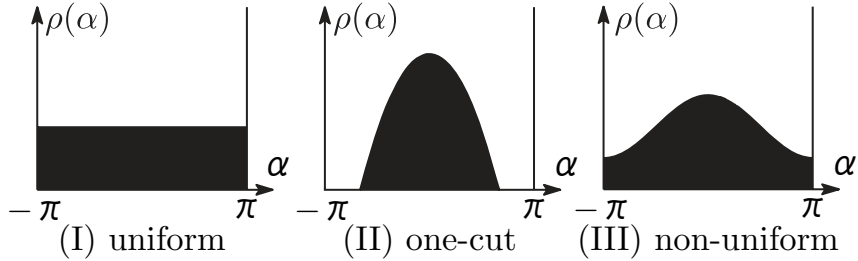}
\caption{The distribution function $\rho (\alpha)$ in three states.}
\label{fig-rho}
\end{center}
\end{figure}

In the finite-$N$ cases, the saddle points are characterised by the positions of the eigenvalues.
At large $N$, the eigenvalue distribution function \cite{Brezin:1977sv}
\begin{align}
\rho (\alpha) \equiv \frac{1}{N} \sum_{i=1}^N \delta(\alpha-\alpha_i)
\end{align}
 is a more convenient tool.
Fig.~\ref{fig-rho} represents its profile, 
and (I) and (II) would correspond to the confinement (\ref{confinement}) and the deconfinement (\ref{deconfinement}) phases, respectively. 
(In (II), we turn on a quantum effect and the eigenvalues compose a cut\footnote{We  use the word ``cut'' for a region which is occupied by the eigenvalues and use the word ``gap'' for an unoccupied region in the distribution function $\rho(\alpha)$
 according to the standard terminology of matrix model \cite{Brezin:1977sv}.}.)

One additional ingredient of the large-$N$ gauge theories is the topology of the eigenvalue distribution function $\rho(\alpha)$.
Although, both (II) and (III) in Fig.~\ref{fig-rho} have one peak, they have different topologies, i.e. (II) has one gap and (III) is gapless.
Thus they must belong to different phases. 
Indeed a third order phase transition related to these two phases occurs in the two-dimensional lattice gauge theory \cite{Gross:1980he, Wadia:1980cp}.
This type of transition is called ``Gross-Witten-Wadia (GWW) transition'' and occurs only at large $N$. 
(The singularity at this transition point is resolved at finite $N$.)
In this way, we need to consider peaks, cuts and gaps in the eigenvalue distribution function in order to classify the phases. 

\begin{figure}
\begin{center}
\includegraphics[scale=1.3]{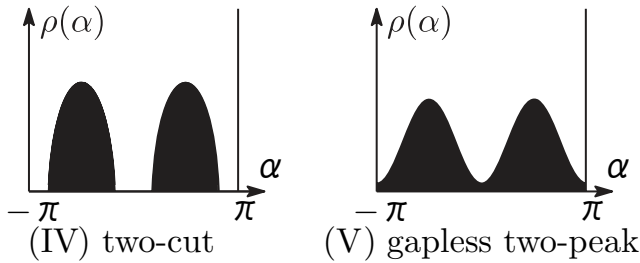}
\caption{The distribution function $\rho(\alpha)$ for two-peak states with two gaps (=two cuts) (IV) and gapless (V).
}
\label{fig-2-cut}
\end{center}
\end{figure}

Now we consider multi-peak solutions in large-$N$ gauge theories.
The basic idea is similar to the finite-$N$ case.
If the deconfinement phase is stable owing to the sufficient attractive forces between the eigenvalues, 
the saddle point solutions may be possible by adjusting the positions.
For example the $Z_2$ symmetric two-cut solution like (IV) in Fig.~\ref{fig-2-cut} may exist.
In this solution, if the attractive forces are not so strong, the cuts  would spread and the gaps might disappear.
Then a GWW type transition might happen and a gapless two-peak solution  like (V) in Fig.~\ref{fig-2-cut} might appear.
In this way, various topologies of multi-peak solutions would exist in the large-$N$ gauge theories.\\

In this section, we have seen the multi-peak saddle point solutions in the gauge theories on $S^1$.
Similar calculations would be possible in weakly coupled $SU(N)$ and $U(N)$ gauge theories, and these solutions may be generally found\footnote{
Although the one-loop effective potential is described by the sum of the potentials between the two eigenvalues like (\ref{qcd-pot}), 
the low energy effective action generally involves higher order couplings like $\alpha_i \alpha_j \alpha_k  \cdots $ in a strong coupling.
Then the intuitive derivation of the saddle points based on the simple attractive force may not work.}.
Note that related multi-peak solutions have been studied in several models by selecting appropriate setups so that these solutions become stable \cite{Unsal:2010qh, Jurkiewicz:1982iz, Mandal:1989ry}.
What we have shown is that these solutions appear not only in these specific models but also in a wide class of gauge theories.

Then a question is what the importance of these solutions is.
We will study their roles in the following sections.

\section{Saddle Point Solutions in the One-dimensional Large-$N$ Gauge Theory}
\label{sec-1d}

In this section, we study the one dimensional gauge theory (\ref{BFSS}) at large $N$.
Although we have studied the multi-peak solutions in the gauge theories in the previous section, the analyses were just perturbative calculations in weak coupling.
We investigate the non-perturbative aspects of the multi-peak solutions in the model (\ref{BFSS}) through a $1/D$ expansion.

\subsection{Phase Structure of the One-dimensional Gauge Theory}
\label{subsec-phase}

Before studying the multi-peak solutions in the one-dimensional large-$N$ gauge theory (\ref{BFSS}), we summarise the phase structure of this model, which has been studied by using a $1/D$ expansion \cite{Mandal:2009vz}.

We take a limit $D,N \to \infty$ and $g\to 0$ with fixed $\tilde{\lambda} \equiv g^2ND$, and then find a non-trivial vacuum, in which $Y^I$ possesses a mass gap $\Delta$, where $\Delta$ satisfies a relation
\begin{align} 
\sum_{I} \langle \Tr Y^I Y^I \rangle  = \frac{N}{2g^2} \Delta^2.
\label{Delta}
\end{align} 
Thus we can integrate out $Y^I$ and obtain an effective action \cite{Mandal:2009vz}: 
\begin{align} 
\frac{S_{\rm eff}(u_n,\Delta)}{DN^2}=-\frac{\beta \Delta^4}{8 \tilde{\lambda}}
+\frac{\beta \Delta}{2}+\sum_{n=1}^{\infty} \frac{1}{n}\left(\frac{1}{D}-e^{-n\beta \Delta} \right)|u_n|^2,
\label{effective-2}
\end{align} 
where $u_n  (=u_{-n}^{*})$ is the generalisation of the Polyakov loop (\ref{Poly-loop}) which is defined by
\begin{align} 
u_n \equiv  \int_{0}^{2\pi} d \alpha~ \rho(\alpha) e^{i n \alpha}  =\frac{1}{N} \sum_{k=1}^N e^{in\alpha_k}.
\label{un}
\end{align}
Hence it satisfies
\begin{align} 
\rho(\alpha)= \frac{1}{2\pi}\left(1 + \sum_{n \neq 0} u_n e^{-in\alpha}  \right).
\label{rho}
\end{align}
The mass gap $\Delta$ is fixed such that this action is extremised.

We evaluate this effective action in order to figure out the phase structure.
Note that $u_n$ for each $n$ can be regarded as independent variables at large $N$, as far as the eigenvalue distribution function $\rho(\alpha)$ does not have any gap.
($\{ u_n\}$ have to satisfy a constraint $\rho(\alpha) \ge 0 $ but it is irrelevant only if $\rho(\alpha)$ is gapless.)
Then we obtain a solution for ${}^\forall \beta$
\begin{align} 
u_n=0 \quad (n \ge1), \quad \quad \Delta = \Delta_0  \equiv \tilde{\lambda}^{1/3}.
\label{un-conf}
\end{align} 
From eq.(\ref{rho}), we obtain $\rho(\alpha)=1/2\pi$ and this solution corresponds to the uniform configuration (\ref{confinement}).
Hence it is a confinement phase.

\begin{figure}
\begin{center}
\includegraphics[scale=.9]{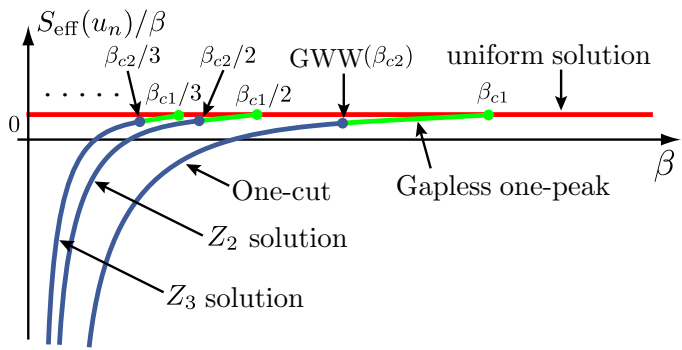}
\caption{A sketch of the effective action $S_{\rm eff}(u_n) $ of the several solutions in the one-dimensional gauge theory (\ref{BFSS}) through the $1/D$ expansion.
The green and blue lines are for the gapless and the gapped solutions respectively.
Note that this sketch is not rigidly plotted based on the equations, and the GWW point $\beta_{c2}$ is much close to $\beta_{c1}$ in the actual plot at large $D$.
}
\label{fig-fene}
\end{center}
\end{figure}

Next we consider a low temperature regime, in which $\exp(-\beta \Delta) \ll 1$ is assumed.
There we can expand the action (\ref{effective-2}) with respect to $\exp(-\beta \Delta)$ and obtain a saddle point for $\Delta$ as
\begin{align} 
\Delta = \tilde{\lambda}^{1/3} \left(1 +\frac{2}{3} \sum_{n=1}^{\infty}e^{- n \beta \tilde{\lambda}^{1/3}} |u_n|^2 \right)+\cdots.
\end{align} 
Then by inserting this solution to the action (\ref{effective-2}),
we obtain an effective action at low temperature
\begin{align} 
&\frac{S_{\rm eff}(u_n)}{DN^2}=\frac{3}{8}\beta \tilde{\lambda}^{1/3}
+\sum_{n=1}^{\infty} \frac{a_n}{n} |u_n|^2 +b_1 |u_1|^4 +\cdots,
\nonumber \\
&a_n=\left(\frac{1}{D}-e^{-n\beta \tilde{\lambda}^{1/3}} \right),
 \quad
b_1=\frac{1}{3} \beta \tilde{\lambda}^{1/3}e^{-2\beta \tilde{\lambda}^{1/3}}.
\label{effective-3}
\end{align} 
This expression implies that the uniform solution (\ref{un-conf}) is  stable if $a_1>0$, which means that the temperature is lower than
\begin{align} 
T_{c1} = \frac{ \tilde{\lambda}^{1/3}}{\log D}+ \cdots.
\end{align} 
Above $T_{c1}$, $a_1$ becomes negative and the solution (\ref{un-conf}) becomes unstable.
On the other hand, from $T_{c1}$, another branch of the solution
\begin{align} 
u_1&=\sqrt{-a_1/2b_1}, \qquad  u_n=0 \quad (n\ge2), \\
\Delta_{1'}& \equiv \tilde{\lambda}^{1/3} \left(1 -\frac{a_1}{3b_1} e^{-  \beta \tilde{\lambda}^{1/3}}  \right)+\cdots
\end{align} 
appears, and this solution is stable above $T_{c1}$.
From eq.(\ref{rho}), the eigenvalue distribution function becomes
\begin{align}
 \rho(\alpha)=\frac{1}{2\pi}\left(1+2 \sqrt{-\frac{a_1}{2b_1}} \cos\left(\alpha \right) \right),
 \label{un-non}
\end{align} 
and this configuration is gapless and one-peak like (III) of Fig.~\ref{fig-rho}.
Note that this solution has broken the shift symmetry of $\alpha$ spontaneously and we have chosen the origin of $\alpha$ such that the peak of $\rho(\alpha)$ is at $\alpha=0$.
In this way, a phase transition between this solution and the uniform solution (\ref{un-conf}) happens at $T_{c1}$, and the order of this transition is second as shown in Ref.~\cite{Mandal:2009vz}.
See Fig.~\ref{fig-fene} too.

The solution (\ref{un-non}) cannot exist above
\begin{align} 
T_{c2} = T_{c1}+ \frac{ \tilde{\lambda}^{1/3}}{6D\log D} + \cdots,
\end{align} 
at which $u_1$ achieves $1/2$ and $\rho(\alpha)$ becomes 0 at $\alpha=\pm \pi$.
A GWW type third order phase transition occurs there and, above this temperature, the configuration possesses one gap like (II) of Fig.~\ref{fig-rho}.
There $u_n$ cannot be regarded as independent variables anymore, and it is difficult to calculate $u_n$ in this regime except $T-T_{c2} \ll \tilde{\lambda}^{1/3}$ as shown in Ref.~\cite{Mandal:2009vz}.
On the other hand, if temperature is sufficiently high $T \gg T_{c2}$,
$A_t$ behaves as a free Gaussian matrix model with a mass $ M \equiv  \sqrt{2D}/\beta \Delta_1$, where
\begin{align} 
\Delta  & = \Delta_1 \equiv \left( 2 \tilde{\lambda}T \right)^{1/4} 
\label{Delta1}
\end{align} 
as we show the details in appendix \ref{app-multi}.
The eigenvalue distribution becomes
\begin{align} 
\rho(\alpha)&= \frac{M^2}{2\pi} {\rm Re}\left(  \sqrt{\frac{4}{M^2}-\alpha^2} \right).
\label{un-gap}
\end{align} 
Here the cut spreads on $[-2/M,2/M]$, and it achieves the $\delta$-functional distribution (\ref{deconfinement}) at $T=\infty$.

Therefore this model has at least three solutions: uniform (\ref{un-conf}), gapless one-peak (\ref{un-non}) and one-gap (= one-cut) (\ref{un-gap}), and these solutions are stable in $[0, T_{c1}]$, $[T_{c1}, T_{c2}]$ and $[T_{c2},\infty]$ respectively.
We can also confirm these stabilities by evaluating the values of the effective action (\ref{effective-2}) at these solutions,
\begin{align} 
\frac{S_{0}}{\beta DN^2} &= \frac{3}{8} \tilde{\lambda}^{1/3},  & (~ {}^\forall T~) ,
\label{F_0}
\\
\frac{S_{1'}}{\beta DN^2}& = \frac{3}{8} \tilde{\lambda}^{1/3} -\frac{a_1^2}{4b_1}T, & (T_{c1} \le T \le T_{c2}) ,
\label{F1'}
 \\
\frac{S_{1}}{\beta DN^2} & =-\frac{3T}{4} \log \left( \frac{T}{\tilde{\lambda}^{1/3}} \right), & ( T_{c2} \ll T).
\label{F1}
\end{align} 
Here the first and second actions are for the uniform and gapless one-peak solutions.
(In this article, we put a prime to the quantities of non-uniform gapless solutions.)
The third one, which is derived from (\ref{energy}), is for the one-cut solution (\ref{un-gap}) at high temperature $T \gg T_{c2}$.
Note that we can evaluate the effective action for the one-cut solution in $T \ge T_{c2}  $ by using a Monte Carlo calculation of the effective action (\ref{effective-2}), and show that $S_1$ is connected to $S_{1'}$ at the GWW point $T_{c2}$.
We sketch these relations in Fig.~\ref{fig-fene}.
From these three actions, we can read off the stabilities.
In $T\ge T_{c1}$, $S_0$ is higher than  $S_{1'}$ or $S_{1}$ and it indicates that the uniform solution is unstable in this region.

Although the $1/D$ expansion predicts the two transitions at $T_{c1}$ and $T_{c2}$, it has not been confirmed in the Monte Carlo calculation of the model (\ref{BFSS}). 
The Monte Carlo just indicates that some transition(s) occurs around a critical temperature $T_{c} \sim 1.30$ at $D=2$ and $T_{c} \sim 0.89$ at $D=9$ \footnote{\label{ftnt-cri}
One reason for the difficulty of the decision of the critical temperature(s) in the Monte Carlo calculation is that $T_{c2}-T_{c1} \sim O(1/D)$ in the $1/D$ expansion and may be small. 
In addition, the transition is not sharp in the  Monte Carlo owing to finite-$N$ effect.
Note that there is a possibility that this transition is a single first-order transition rather than the two transitions \cite{Aharony:2004ig,  Aharony:2005ew, Aharony:2003sx, AlvarezGaume:2005fv}.
We will fix this problem in a future work.}.
(Here we have used the unit $\lambda \equiv g^2N=1$, which we will use throughout the numerical calculation in this article.)
However the details of the properties around $T_c$ might not be crucial in our study.

\subsection{$Z_m$ Symmetric Saddle Point Solutions}
\label{sec-Zm}

Now we derive the multi-peak saddle point solutions from the effective action (\ref{effective-2}).
At high temperature $T \gg T_{c2}$ , we can numerically derive an $m$-cut solution for a given set of the numbers of the eigenvalue $\{N_1,N_2,\cdots ,N_m \}$ for each cut, if $N_l \sim O(N)$ and $m \ll N$, as shown in appendix \ref{app-multi}.
In this section, instead of considering this generic situation, we study the $Z_m$ symmetric situation, which is a generalisation of the configuration (\ref{Zm}).
In this case, the analysis is much simpler and we can derive $Z_m$ solutions analytically.

Recall that $u_n$ can be treated as independent variables in the effective action (\ref{effective-2}).
Thus we have a classical solution for ${}^\forall \beta $,
\begin{align} 
u_n=0, \quad (n \neq km, ~~k=1,2,\cdots)  .
\end{align} 
We substitute this solution into the action (\ref{effective-2}) and obtain an effective action for $\Delta$ and $u_{km}$ $(k=1,2,\cdots)$,
\begin{align} 
\frac{S_{\rm eff}(u_{km},\Delta)}{DN^2}
= \frac{1}{m} \Biggl[& -\frac{ m\beta \Delta^4}{8 \tilde{\lambda}}
+\frac{m \beta \Delta}{2} 
+\sum_{k=1}^{\infty} \frac{1}{k}\left(\frac{1}{D}-e^{-km\beta \Delta} \right)|u_{km}|^2 
\Biggr].
\label{effective-m}
\end{align} 
This action is the same expression as the original action (\ref{effective-2}) by replacing $u_n \to u_{nm}$ and  $ \beta \to m \beta $,  except the overall factor $1/m$.
Therefore if $\{ u^{(cl)}_n \} $ $(n=1,2,\cdots)$ is a solution of the action (\ref{effective-2}) at an inverse temperature $\beta$, $u_{nm} = u_n^{(cl)}$ $(n=1,2,\cdots)$ is a solution of the new action (\ref{effective-m}) at the inverse temperature $m\beta$.

By using this trick, we obtain the $Z_m$ symmetric gapless $m$-peak solution from (\ref{un-non}),
\begin{align} 
u_m=&\sqrt{-a_m/2b_m}, \qquad u_n=0 \quad (n \neq m) ,
\label{sol-Zm}
\\
\Delta=&\Delta_{Z_m'} \equiv \tilde{\lambda}^{1/3} \left(1 -\frac{a_m}{3b_m} e^{-  m\beta \tilde{\lambda}^{1/3}}  \right), \\
b_m=&\frac{1}{3}m \beta \tilde{\lambda}^{1/3}e^{-2m\beta \tilde{\lambda}^{1/3}}, \nonumber
\end{align} 
where $a_m$ has been defined in eq.(\ref{effective-3}).
From eq.(\ref{rho}), the eigenvalue density becomes
\begin{align}
 \rho(\alpha)=&\frac{1}{2\pi}\left(1+2 \sqrt{-\frac{a_m}{2b_m}} \cos\left( m \alpha \right) \right) .
\label{rho-Zm}
 \end{align}
 Indeed this configuration has a $Z_m$ symmetry: $\alpha \to \alpha +2\pi/m$.
 The above arguments imply that  this solution exists in $mT_{c1} \le T \le  mT_{c2} $.
At $T=mT_{c2}$, a GWW transition happens and $m$ gaps arise.
Then this solution develops to the $Z_m$ symmetric $m$-cut solution, which is given by (\ref{rho-m}) with $N_l=N/m$ at high temperature $T \gg mT_{c2}$.
In this $m$-cut solution, $u_{km} \neq 0$ and $u_n=0$ for $n\neq km $  
 ($k=1,2,\cdots$) are satisfied.
 
Similarly we obtain the observables in $Z_m$ solution.
From eq.(\ref{F1'}), the value of the effective potential for the gapless $m$-peak solution (\ref{sol-Zm}) at $mT_{c1} \le T \le  mT_{c2} $ is given by
\begin{align} 
\frac{S_{Z_m'}}{\beta DN^2}& =  \frac{3}{8} \tilde{\lambda}^{1/3} -\frac{a_m^2}{4mb_m}T.
\end{align} 
At high temperature $T \gg m T_{c2}$, $\Delta$ and the effective action of the $Z_m$ symmetric $m$-cut solution becomes
\begin{align} 
\Delta_{Z_m} & =  \left(  2 \tilde{\lambda}T/m \right)^{1/4} , \qquad 
\frac{S_{Z_m}}{\beta DN^2}  = 
-\frac{3T}{4m} \log \left(  \frac{T}{m \tilde{\lambda}^{1/3}} \right)
\label{energy-m}
\end{align} 
from eq.(\ref{Delta1}) and (\ref{F1}).
Hence $Z_m$ solution for larger $m$ has a higher potential and is more unstable.
See Fig.~\ref{fig-fene}.
We will compare these results with the Monte Carlo calculation in Sec.~\ref{subsec-MC}.

Lastly we argue the stability of $Z_m$ solution.
In the effective action (\ref{effective-m}), $Z_m$ solution is the stable solution for $T>mT_{c1}$.
It indicates that $Z_m$ solution is stable against the perturbation with respect to $u_{nm}$.
On the other hand, $a_n$ ($1\le n\le m-1$) are negative at $T>mT_{c1}$, and $u_n=0$ ($1\le n\le m-1$) is unstable in the original action (\ref{effective-2}).
Thus $Z_m$ solution is unstable against the perturbation with respect to $u_n$   ($1\le n\le m-1$) and is a saddle point as we expected.

\section{Stochastic Time Evolution in the One-dimensional Gauge Theory}
\label{sec-cascade}

We have shown that the model (\ref{BFSS}) has many saddle point solutions.
However we have still not fully revealed the natures of the model (\ref{BFSS}).
One missing property is the profile of the effective potential for $\{ u_n \}$ (or $\{ \alpha_i \}$), which we have seen in the $SU(3)$ case in Fig.~\ref{fig-SU3}.
Such a profile is crucial to investigate the decay process of an unstable state to a stable state.
Although our effective potential is thermal and is not directly related to the actual real-time decay process,  if the decay happens sufficiently slowly or adiabatically, the decay process may reflect the profile of the potential.

Of course we cannot draw the potential like the $SU(3)$ case, since we have infinite variable $\alpha_i$ at large $N$.
However we possibly read off the relevant part of the profile of the potential through the following procedure.
Suppose that we can integrate out adjoint scalar $Y^I$ in the model (\ref{BFSS}) and obtain an effective action for the gauge field  $S_{{\rm eff}}(\alpha_i)$.
Then we consider an unstable solution in the action $S_{{\rm eff}}(\alpha_i)$. 
(It corresponds to $\blacktriangle$ in Fig.~\ref{fig-SU3}.)
We smoothly deform $\{ \alpha_i \}$ from this solution such that $S_{{\rm eff}}(\alpha_i)$ is becoming smaller.
By repeating this deformation, $\{ \alpha_i \}$ may settle down to a stable configuration finally. (It may correspond to $\blacksquare$ in Fig.~\ref{fig-SU3}.)
Then we can speculate the profile of the potential between the unstable solution and the stable configuration
from the history of the deformation of $\{ \alpha_i \}$.

We investigate this process by using a stochastic time evolution of a Monte Carlo calculation, which we will explain in the next subsection.
Then we will see that this method captures several characteristic properties of the model (\ref{BFSS}), which are related to the dual gravity as we will argue in Sec.~\ref{sec-gravity}.

\subsection{Overview of the Stochastic Time Evolution}
\label{subsec-stochastic}

We introduce the stochastic time evolution, which is designed such that the above process is realised.
We assign a discrete ``stochastic time'' $s$ for $\alpha_i$ and $Y^I$, which is distinguished from the Euclidean time $t$ in the model (\ref{BFSS}).
We take an appropriate initial configuration at $s=0$, and,
to gain the time $s$, we update $\alpha_i(s)$ and $Y^I(s)$ through the following rule:
\begin{enumerate}
  \item Set a trial configuration $\alpha_{i,{\rm trial}}(s+1) = \alpha_i(s)+r$, where $r$ is a small random number.
  \item If $S[\alpha_{i,{\rm trial}}(s+1)] \le S[\alpha_i(s)]$, we accept this trial configuration as $\alpha_{i}(s+1)= \alpha_{i,{\rm trial}}(s+1) $.
\item Even if $S[\alpha_{i,{\rm trial}}(s+1)] > S[\alpha_i(s)]$, we still accept  $\alpha_{i,{\rm trial}}(s+1) $  with the probability $ \exp (-S[\alpha_{i,{\rm trial}}(s+1)] + S[\alpha_i(s)] )$, and, if it is rejected, we retain $\alpha_{i}(s+1)= \alpha_i(s)$.
\item Update the scalar field  $Y^{I}(s)$ sufficiently many
 times\footnote{
The number of updating $Y^{I}$ in each $s$ is fixed and
the detailed balance condition is satisfied.
See the details in appendix B.}
 such that they arrive at an equilibrium for the given configuration $\{\alpha_i(s+1) \}$, and use this state as $Y^I(s+1)$.
\end{enumerate}
The last step might correspond to the path integral of $Y^I$ and is taken to focus on the dynamics of $\{ \alpha_i \}$ in  $S_{{\rm eff}}(\alpha_i)$. 
Through this evolution, $\{ \alpha_i \}$ is deformed gradually such that 
the action $S_{{\rm eff}}(\alpha_i)$ tends to be smaller as we intended.
Note that this procedure for $\{ \alpha_i \}$ is based on the Metropolis algorithm.

The technical details of this Monte Carlo  calculation are explained in appendix \ref{app-MC}.

\subsection{Decay Patterns in the Stochastic Time Evolution}
\label{subsec-decay}

\begin{figure}
\begin{center}
\includegraphics[scale=0.55]{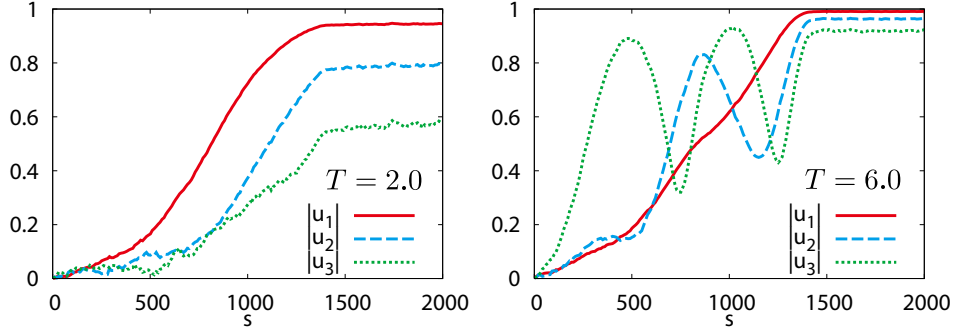}
\end{center}
\caption{Two decay patterns of the stochastic time evolution of the uniform solution (\ref{un-conf}).
$s$ is the stochastic time, and we take $D=9$ and $N=60$.
(LEFT) The direct decay observed at $T=2.0 (\sim 2.2T_c)$. (RIGHT)
The cascade decay at $T=6.0 (\sim 6.7T_c)$.
} 
\label{fig-un}
\end{figure}

We investigate the stochastic time evolution of the unstable states of the model (\ref{BFSS}).
At $T>T_{c}$   \footnote{$T_c \sim 1.3$ at $D=2$ and $T_c \sim 0.89$ at $D=9$ in the Monte Carlo  calculation of the model (\ref{BFSS}) as we mentioned at the end of Sec.~\ref{subsec-phase}.}, we take the unstable uniform solution (\ref{un-conf}) (confinement configuration) as the initial state, and evaluate the evolutions repeatedly by changing the temperature and random number.
Then we observe the following two evolution patterns depending on temperature\footnote{The movies for the stochastic evolutions are available on
\\ \url{http://www2.yukawa.kyoto-u.ac.jp/~azuma/multi_cut/index.html}.}:

\paragraph{Direct Decay}
In this pattern, the unstable uniform solution directly evolves to a more stable one-peak state (deconfinement configuration).
See the left plot of Fig.~\ref{fig-un}.
In this figure, we plot the stochastic time evolution of $|u_n|$ with respect to the stochastic time $s$.
Through eq.(\ref{un}), $|u_n|$ roughly indicates the number of the peaks of the eigenvalue density $\rho(\alpha)$.
If $m$ peaks exist in  $\rho(\alpha)$, $u_m$ tends to have a stronger signal relative to other $u_n$ ($n\neq m$).
If all $u_n$ are close to zero, it indicates an uniform distribution.
Indeed, in Fig.~\ref{fig-un} (LEFT), $|u_n| \sim 0$ around $s=0$ indicates the uniform state (\ref{un-conf}) and 
the signal $|u_1|>|u_2|>\cdots$ in the late history indicates the one-peak state.
This pattern is mainly observed at lower temperature $T_{c}<T < c(D)T_{c}$.
Here $c(D)$ is a constant, which seems to depend on $D$ as $c(D=2)\sim 3.5$ and $c(D=9)\sim 5.0 $, although the change of the decay pattern at $c(D)T_{c}$ is not sharp.

\paragraph{Cascade Decay}
In this pattern, the unstable uniform solution first evolves to a multi-peak state.
Then the peaks attract each other, and two of them collide and merge into one peak.
By repeating such collisions, the number of the peaks decreases one by one, and it finally achieves the one-peak state.
This pattern is dominant at high temperature $c(D)T_{c} <T$.
The $m$-peak states with a larger $m$ tend to appear at higher temperature.

\begin{figure}
\begin{tabular}{cccc}
\begin{minipage}{0.25\hsize}
\begin{center}
\includegraphics[width=3cm]{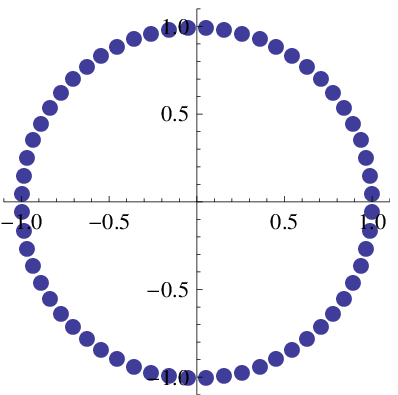}
$s=0$
\end{center}
\end{minipage}
\begin{minipage}{0.25\hsize}
\begin{center}
\includegraphics[width=3cm]{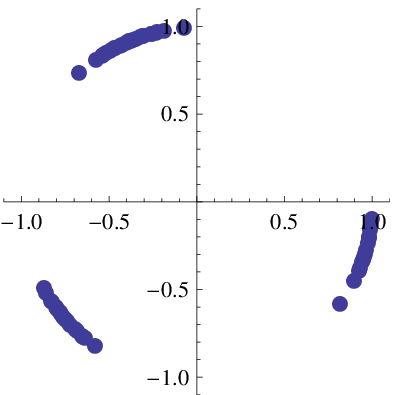}
$s=500$
\end{center}
\end{minipage}
\begin{minipage}{0.25\hsize}
\begin{center}
\includegraphics[width=3cm]{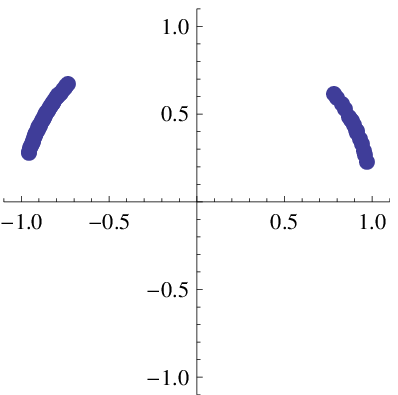}
$s=1000$
\end{center}
\end{minipage}
\begin{minipage}{0.25\hsize}
\begin{center}
\includegraphics[width=3cm]{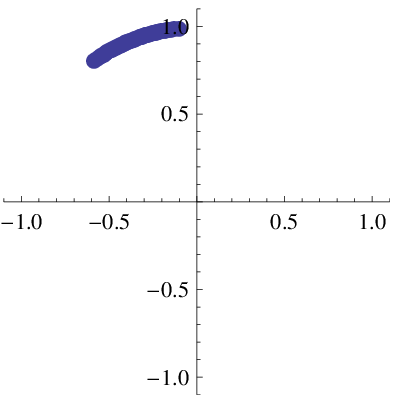}
$s=1500$
\end{center}
\end{minipage}
\end{tabular}
\caption{
The stochastic evolution of the eigenvalue distribution in the right plot (the cascade decay) of Fig.~\ref{fig-un}.
$e^{i \alpha_k}$ ($k=1,\cdots, N$) are plotted.}
\label{fig-cascade}
\end{figure}

One example is shown in the right plot of Fig.~\ref{fig-un} and Fig.~\ref{fig-cascade}.
In this example, a three-peak state  as shown in Fig.~\ref{fig-cascade} appears around $s=300$.
Correspondingly a strong signal appears at $|u_3|$ in Fig.~\ref{fig-un}.
Then the two peaks merge into one and it becomes a two-peak state around $s=800$, and finally they collapse to a one-peak state around $s=1300$.
In Fig.~\ref{fig-un}, the strong signal of $|u_2|$ around $s=800$ corresponds to the two-peak state but the strong signal of $|u_3|$ around $s=1000$ is merely owing to the asymmetry of the eigenvalue distribution during the decay.\\

In this way, the multi-peak states appear as the intermediate states in the stochastic decay process of the unstable uniform state\footnote{We observe the similar decay patterns in the Monte Carlo calculation of the effective action (\ref{effective-2}) too but we omit to show the results here. 
In this calculation, we have to find $\Delta$, at which the action (\ref{effective-2}) becomes the maximum rather than the minimum, since $\Delta$ is originally a complex variable \cite{Mandal:2009vz}. }. 
These results indicate that the multi-peak states lie between the uniform and the one-peak configuration in the potential valley of $S_{{\rm eff}}(\alpha_i)$.
Note that, although the stochastic time evolution is powerful to capture these qualitative features of the potential, we have not established a proper method to investigate them quantitatively.

\begin{figure}
\begin{center}
\includegraphics[scale=0.55]{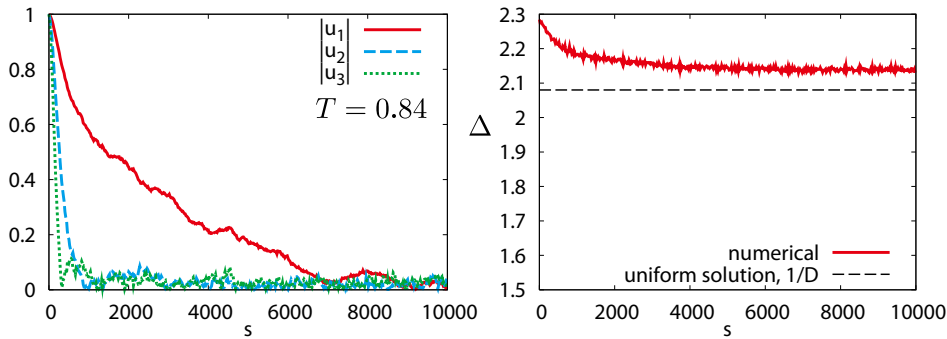}
\end{center}
\caption{The decay of an unstable one-peak configuration to an uniform state at $T=0.84 (\sim 0.9 T_c) $ with $N=60$ and $D=9$.
In the right, we compare $\Delta$ with the $1/D$ expansion for the uniform solution (\ref{un-conf}).
}
\label{fig-one}
\end{figure}

Similarly we attempt the stochastic time evolution of an unstable one-peak configuration at low temperature $T<T_{c}$, but it merely evolves to the uniform configuration as shown in Fig.~\ref{fig-one}.
This would be consistent with the fact that the uniform solution (\ref{un-conf}) may be an unique solution at $T<T_{c}$ \footnote{
One issue is around $T_c$.
As we mentioned in the footnote \ref{ftnt-cri}, the order of the phase transition of the model (\ref{BFSS}) has not been fixed.
If the transition is the first-order, meta-stable solutions must exist around $T_c$  \cite{Aharony:2004ig,  Aharony:2005ew, Aharony:2003sx, AlvarezGaume:2005fv}. 
}.\\

Note that we have to take care of the topology of the eigenvalue density $\rho(\alpha)$ in the Monte Carlo calculation.
If we smoothly deform $\rho(\alpha)$ at large $N$, its topology never changes. 
The topology change must accompany some singularity such as the GWW transition.
Thus if we take the uniform configuration as the initial state, any gap never appears during the evolution.

On the other hand, the Monte Carlo calculation is done at finite $N$, and we cannot define the topology of $\rho(\alpha)$ rigidly. 
Therefore we have to speculate the topology at large $N$ from the result of the Monte Carlo calculation.
For example, no gap should appear in the distribution function $\rho(\alpha)$
in the stochastic time evolution starting with the uniform
configuration, if it were in large $N$.
Hence we should interpret that the multi (one) peaks of $\rho(\alpha)$ in Fig.~\ref{fig-cascade} are connected by tiny non-zero eigenvalue densities.
Since such tiny densities cost little energy, they may be irrelevant in the dynamics.
However they may be relevant in the gauge/gravity correspondence, since the singularity associated with the topology change may correspond to the naked singularity in the gravity \cite{BMMW, AlvarezGaume:2005fv,  Liu:2004vy,  AlvarezGaume:2006jg}.

\subsection{Comparison of the Monte Carlo with the $1/D$ Expansion about $Z_m$ Solution}
\label{subsec-MC}

\begin{figure}
\begin{center}
\includegraphics[scale=0.55]{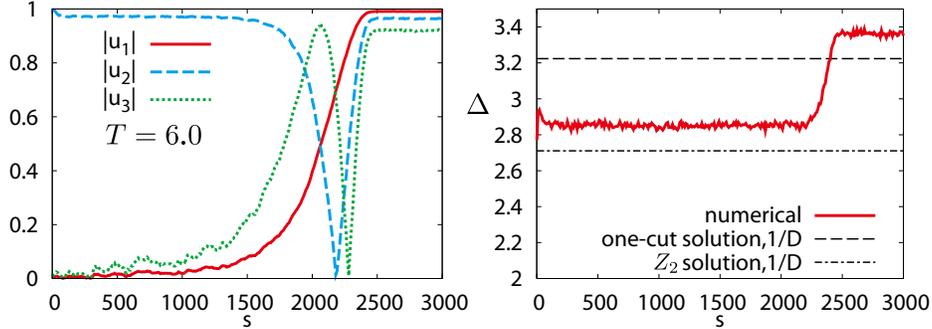}
\end{center}
\caption{
The stochastic time evolution of a $Z_2$ symmetric state.
We take $D=9$, $N=60$ and $T=6.0 (\sim 6.7 T_c)$.
$Z_2$ solution may be realised between $s=100$ and $s=1000$.
At that period, the observables are also stable as shown in the right plot.
We evaluate the averages of the observables there and compare them with the results from the $1/D$ expansion as summarised in Table \ref{table-Zm}.
Note that $\Delta$ in the right plot remains almost constant until around $s=2200$, although $u_n$ start evolving around $s=1000$.
We will discuss this reason in appendix \ref{app-multi}. 
}
\label{fig-ZmMC}
\end{figure}

We have studied the decay patterns in the stochastic time evolutions through the Monte Carlo (MC) calculation.
By applying this method, we evaluate the observables of $Z_m$ solution, and compare the results from the $1/D$ expansion.

To obtain a $Z_m$ solution in the MC calculation, we take an appropriate $Z_m$ symmetric configuration as the initial state of the stochastic time evolution.
We evaluate this evolution and find that a $Z_m$ solution may be realised as a long life state\footnote{The terminology ``long life'' would not be proper, since this state is not meta-stable.} at the initial stage of this process at High temperature\footnote{We tried to observe any signal for the critical temperature for $Z_m$ solution around $mT_{c}$. However, the life time of the $Z_m$ solution around $mT_{c}$ seemed too short and we have not found any clear evidence. }.
The solution finally decays to the one-peak configuration, which is stable at $T > T_{c}$.
See Fig.~\ref{fig-ZmMC} for the $Z_2$ case.
From the arguments below eq.(\ref{rho-Zm}), $Z_2$ solution has a strong signal at $|u_2|$, and
 $|u_{2n+1}|$ should be zero. 
It indeed happens in Fig.~\ref{fig-ZmMC} until around $s=1000$.
Thus we believe that this is the $Z_2$ symmetric two-cut solution.
Then we compare the observables of the MC and the $1/D$ expansion, and these are consistent as summarised in Table \ref{table-Zm}.

\begin{table}
\begin{tabular}{l|cccccc}
\hline
& $\Delta_{Z_1}$ & $\Delta_{Z_2}$  & $\Delta_{Z_3}$ & $E_{Z_1}/N^2$  & $E_{Z_2}/N^2$ & $E_{Z_3}/N^2$  \\ \hline
MC & 3.3625(5) & 2.850(2) & 2.602(2) &  36.71(2) & 19.14(4) & 13.46(5) \\ 
$1/D$ &  3.22  &  2.71 & 2.45 &  40.5 & 20.3 & 13.5 \\
\hline
\end{tabular}
\caption{The comparison between the $1/D$ expansion and  Monte Carlo calculation of the model (\ref{BFSS}) for $Z_m$ solution. 
$N$, $D$ and $T$ are the same as the data in Fig.~\ref{fig-ZmMC}.
We use the unit $\lambda=1$.
We obtain $Z_3$ solution by a similar fashion to $Z_2$ solution.
We evaluate $\Delta$ defined in (\ref{Delta}) and the internal energy  
$E \equiv -\frac{d}{d \beta} \log Z $ \cite{Kawahara:2007fn}. 
The $1/D$ expansion predicts $\Delta_{Z_m}$ and
$E_{Z_m}/DN^2 = 3T/(4m)+O(1/D)$ from eq.(\ref{energy-m}).
}
\label{table-Zm}
\end{table}

\section{Multi-cut Solutions in the Dual Gravity}
\label{sec-gravity}

\subsection{The Dual Supergravity of the One-dimensional Gauge Theory}
\label{subsec-dual-gr}

We have studied  the multi-peak solutions in the gauge theory (\ref{BFSS}).
Now we consider their dual geometries through  the gauge/gravity correspondence \cite{Maldacena:1997re, Itzhaki:1998dd}.
In this subsection, we review the dual gravity of the model  (\ref{BFSS}) based on the arguments in Ref.~\cite{Aharony:2004ig}.
Roughly speaking, the eigenvalue distribution of $\{ \alpha_i \}$ is related to the positions of black D0 branes on the spatial circle in a Kaluza-Klein gravity.

We consider IIB string theory on $R^8 \times S^1_{\beta_2} \times S^1_L $, where we have defined the inverse temperature as $\beta_2$ and the period of the spatial circle as $L$.
We fix the periodicity of the fermions along $S^1_L$ as periodic.
We put $N$ D$1$ branes, which wind this $S^1_{\beta_2} \times S^1_L $, and 
then the effective theory of the branes is given by the two-dimensional maximally supersymmetric $U(N)$ Yang-Mills theory on the $ S^1_{\beta_2} \times S^1_L$.
\begin{align} 
S_{{\rm 2dSYM }}=&\frac{1}{g_2^2} \int_0^{\beta_2} \! dt  \int_0^L \! dx^1 
 \Tr \Biggl[ \frac{1}{2} F_{01}^2 +\sum_{I=2}^9 \frac{1}{2} \left( D_\mu Y^I \right)^2
-\frac{1}{4}\sum_{I,J} [Y^I,Y^J]^2+ {\rm fermions}
\Biggr].
\label{2dSYM}
\end{align} 
In this theory, the temporal Kaluza-Klein (KK) modes and fermions are decoupled, if the temperature is sufficiently high $\beta_2 \ll (L/\lambda_2)^{1/3} $, where $\lambda_2=g^2_2N$.
Then the theory is reduced to the one-dimensional model~(\ref{BFSS}) with $D=9$ and $\lambda = \lambda_2/\beta_2$.
Here we have to identify 
\begin{align}
\begin{tabular}{ccc}
2d SYM on  $S^1_{\beta_2} \times S^1_L$ &  & 1d model on $S^1_{\beta} $ (\ref{BFSS})  \\ 
 $x^1$, $L$ & $\leftrightarrow$   &  $t$, $\beta$, \\
  $A^{(0)}_{x^1}$ & $\leftrightarrow$  &  $A_t$, \\
  $A^{(0)}_t$ & $\leftrightarrow$ & $g Y$,
  \end{tabular}
\label{KK-red}
\end{align}
where ``$(0)$'' in the two dimensional fields denote the zero modes with respect to the temporal KK expansion.

Through the gauge/gravity correspondence, the gauge theory (\ref{2dSYM}) at large $N$ can be described by the black D1 brane geometry in the IIB supergravity \cite{Itzhaki:1998dd}
\begin{align} 
&\frac{ds^2}{\alpha'}=
F^{-1/2}\left(  fdt^2 +dx_1^2 \right) +F^{1/2}\left( \frac{du^2}{f} +u^2 d \Omega^2_7
\right) , \nonumber \\
&F = \frac{2^6 \pi^3 \lambda_2}{u^6}, \quad f= 1-\frac{u_0^6}{u^6},
\quad u_0^2=\frac{16\pi^{\frac{5}{2}} }{3 } \frac{\sqrt{\lambda_2}}{\beta_2}
\label{metric-D1}
\end{align} 
with a dilaton and Ramond-Ramond (RR) potential.
This gravity description is valid at strong coupling $\lambda_2 \beta_2^2 \gg 1$ so that the $\alpha'$ corrections are suppressed.
In addition to this condition, $L$ should be larger than the effective string length $\beta_2^{3/4}/\lambda_2^{1/8} $ at the horizon so that the excitations of the winding strings along $S^1_L$ are suppressed.
However interesting phenomena will occur when $L$ is smaller than this effective string length.
Although the IIB supergravity does not work in this regime, we can avoid this problem by taking a T-duality on the  $S^1_L$ and go to the IIA frame.
Here the original IIB and IIA theory are related as
\begin{align}
\begin{tabular}{ccc}
IIB  &  T-dual on $S^1_L$  & IIA  \\ 
$S^1_L$ &$\leftrightarrow$ & $S^1_{L'}$~ ($L' \equiv (2\pi)^2/L$) , \\
D1 brane  winding $S^1_L$ & $\leftrightarrow$ & D0 brane localised on $S^1_{L'}$ \\
$A_{x^1}$   &$\leftrightarrow$ & $Y^1$  .
\end{tabular}
\label{T-dual}
\end{align}
Thus the period of the dual circle, which we have defined as $L'=(2\pi)^2/L$, can be large for a small $L$, and the IIA supergravity description works if $L \ll \beta_2^{3/4}/\lambda_2^{1/8} $.
Under this T-duality, the black D1 brane (\ref{metric-D1}) is mapped to
\begin{align} 
&\frac{ds^2}{\alpha'}=
F^{-\frac12} fdt^2+F^{\frac12}\left( \frac{du^2}{f}+d{x'_1}^{2} +u^2 d \Omega^2_7
\right),
\label{metric-D0}
\end{align}
where $x_1'=x_1'+L'$. 
This geometry represents that the $N$ D0 branes, which are the T-dual of the D1 branes as in (\ref{T-dual}), are  uniformly distributed along $S^1_{L'}$, and is called ``smeared black D0 brane''.
Note that the topology of the horizon of this geometry is $S^1_{L'}\times S^7$, and this geometry is a kind of a black string in  Kaluza-Klein gravity.

However this geometry is stable only for small $L'$ \cite{Aharony:2004ig, Harmark:2004ws,Barbon:1998cr}.
Around $L' \sim \lambda_2^{1/4}/\beta_2^{1/2}$, owing to the Gregory-Laflamme instability \cite{Gregory:1994bj}, the smeared black D0 brane becomes unstable. 
A first order phase transition occurs and the stable geometry for larger $L'$ is given by a black D0 brane localised on a point on $S^1_{L'}$.
In this geometry, the horizon is also localised on $S^1_{L'}$ and its topology is $S^8$.
This geometry is called ``localised black D0 brane'' and the metric is approximately derived in Ref.~\cite{Harmark:2004ws}.

Let us compare this gravity result with the one-dimensional gauge theory (\ref{BFSS}). 
Note that the gravity is valid at low temperature $ \beta_2 \gg 1/\lambda_2^{1/2}$, whereas the gauge theory description is valid at high temperature $\beta_2 \ll (L/\lambda_2)^{1/3}$, and these two parameter regions are different.
Nevertheless we can see that the stability of the gravity is qualitatively consistent with that of the gauge theory studied in Sec.~\ref{subsec-phase}.

The T-duality (\ref{T-dual}) maps the gauge field $A_{x^1}$ on the D1 branes to the adjoint scalar $Y^1$, which represents the positions of the D0 branes on $S^1_{L'}$.
Hence the eigenvalue distribution of $A_{x^1}$ may
correspond to the D0 brane distribution on $S^1_{L'}$.
Through the relation (\ref{KK-red}), the uniform configuration (\ref{un-conf}) and the one-cut configuration like (\ref{un-gap}) may be identified with the smeared black D0 brane (\ref{metric-D0}) and localised black D0 brane respectively.
Recall that the uniform configuration is stable for large $L$, which corresponds to a large $\beta$ in the model (\ref{BFSS}), and the one-cut one is stable for small $L$.
Through the relation  $L'=(2\pi)^2/L$, this stability is consistent with the IIA supergravity\footnote{The order of the phase transition is different. In the gravity, the first order transition is predicted \cite{Aharony:2004ig}.
On the other hand, the gauge theory predicts the two higher order transitions at $T_{c1}$ and $T_{c2}$ through the $1/D$ expansion \cite{Mandal:2009vz}.
This is not inconsistent, since the order of the transition generally  depends on the couplings.
Besides the Monte Carlo calculation of the model (\ref{BFSS}) has not fixed the order of the transition at $T_c$.
}.

\subsection{Gravity Duals of the Multi-cut Solutions}
\label{subsec-dual-m-cut}

Now we consider the dual geometries of the $m$-cut solutions in the gauge theory (\ref{BFSS}).
Through the correspondence between the eigenvalue distribution of $A_{x^1}$ and the distribution of the D0 branes on $S^1_{L'}$, the $m$-cut solutions would correspond to $m$ black D0 branes localised on $S^1_{L'}$.
Such multi black branes have been known as the solutions of the supergravities, and
 are unstable due to the gravitational attractive forces between the branes.
 It is consistent with the instability of multi-cut solutions in the gauge theory.
Indeed the existence of the dual phases in the gauge theories corresponding to these multi black brane solutions in the Kaluza-Klein gravities has been predicted in Ref.~\cite{Harmark:2004ws}.
The saddle point solutions in our article provide evidence for this conjecture.

\subsection{Black String Decay}
\label{subsec-bs}

\begin{figure}
\begin{center}
\includegraphics[scale=.75]{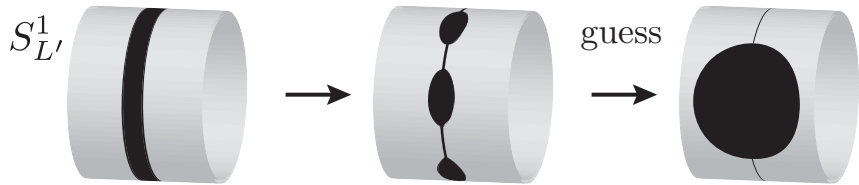}
\caption{A sketch of the decay process of the meta-stable black string in  Refs.~\cite{Choptuik:2003qd, Lehner:2010pn}.
The last step  is just a speculation.
The inside of the apparent horizon is depicted as the black region.
In Ref.~\cite{Lehner:2010pn}, an interesting fractal structure, which is akin to a low viscosity fluid, is observed but we omit it in this sketch.
}
\label{fig-GL}
\end{center}
\end{figure}

Interestingly such multi black holes (connected by thin black strings) appear in a real-time decay process of a black string \cite{Choptuik:2003qd, Lehner:2010pn, Lehner:2011wc}.
As we mentioned, the black string is unstable at large $L'$, and a localised black hole is thermodynamically favoured.
The authors in Refs.~\cite{Choptuik:2003qd, Lehner:2010pn} examine the decay process of a meta-stable black string\footnote{In the model in Refs.~\cite{Choptuik:2003qd, Lehner:2010pn}, the Gregory-Laflamme transition is first order and a meta-stable black string solution exist around the critical point. } by solving the Einstein equation numerically. 
They found that several points of the horizon are growing and other parts are shrinking owing to the Gregory-Laflamme instability.
As a result, a sequence of black holes joined by thin black string segments appear along $S^1_{L'}$.
See Fig.~\ref{fig-GL}.
However the numerical analysis did not work for sufficiently long time and the final state of this process has not been found.
Since this configuration is unstable, it might evolve to a single localised black hole joined by an extremely thin black string.

Remarkably this time evolution of the black string is similar to the stochastic time evolution of the unstable uniform configuration in the gauge theory.
The multi-peak state appears as the intermediate state in both cases.
This agreement may imply that the dynamical stability in the gravity may be related to the thermodynamical stability in the gauge theory\footnote{In gravity, the connection between the dynamical stability and thermodynamical stability is argued in Ref.~\cite{Hollands:2012sf}.}.
(Recall that the stochastic time evolution may reflect the profile of the thermodynamical effective potential of the gauge theory.)

Note that, it is expected that the thin black string segment appearing in the above decay process would be pinched off through a quantum effect, when the size of the horizon achieves the Plank scale.
Although it is difficult to investigate such quantum effect in the gravity, it may be possible in the gauge theory, in which the quantum effect may correspond to the finite-$N$ effect.
Hence it may be important to study the model (\ref{BFSS}) further to uncover this problem. 

However there are several discrepancies between the black string decay \cite{Choptuik:2003qd, Lehner:2010pn} and our stochastic evolution in the model (\ref{BFSS}).
We will discuss them in Sec.~\ref{sec-conclusion}.

\section{Conclusions}
\label{sec-conclusion}

We have found the multi-peak saddle point solutions in various gauge theories on $S^1$.
The important condition for the existence of these solutions is the attractive forces between the eigenvalues, which typically realise in case the deconfinement phase is stable.
Since many maximally supersymmetric Yang-Mills theories are deconfined at any finite temperature, these solutions may exist there.
Then their gravity duals may also exist in the supergravities and
they might belong to a new class of geometries.
To show it, we are studying the supersymmetric gauge theories and will report on a future work.

Through the stochastic time evolution of the one-dimensional gauge theory (\ref{BFSS}), we found that these multi-peak states may exist as the intermediate states between the unstable confinement phase and the stable deconfinement phase at high temperature.
This result indicates that, even in  the real-time decay process of the unstable confinement phase, some related intermediate states might appear.
It is valuable to investigate it further but one difficulty is the definition of the multi-peak states in the real-time formalism, since the theory does not have the temporal circle anymore.

A natural generalisation of this study is the stochastic evolution of an unstable confinement phase in higher dimensional gauge theories.
In this case, $\alpha_i$ can depend on the spatial positions, and this may lead to much richer structures involving the domain walls.

We can also take the $S^1$ as the spatial circle as we considered in Sec.~\ref{sec-gravity}. 
Then we do not have any difficulty about the definition of the multi-peak solutions in the real-time formalism.
One interesting application is to the phenomenology of particle physics.
In some models, the gauge field of the spatial $S^1$, which is an extra-dimension of our world, is identified as the Higgs field  \cite{Manton:1979kb, Fairlie:1979at, Hosotani:1983xw}.
If we consider the time evolution of such models in early universe, 
certain multi-peak configurations might appear and probably play some roles.\\

We have also studied the application to the gauge/gravity correspondence  by identifying the eigenvalue distribution of the spatial gauge field and the positions of the D branes on the spatial circle in the Kaluza-Klein gravity through the T-duality.
There the multi-peak solutions in the gauge theory may correspond to the multi black branes localised on the circle\footnote{
In the case of the bosonic gauge theories like the model (\ref{BFSS}) and pure Yang-Mills theories in the Euclidean signature, there is no distinction between the temporal circle and the spatial one.
Hence we can construct the gravity duals of the multi-cut solutions along the temporal circle too based on the proposal in Ref.~\cite{Witten:1998zw} and \cite{Mandal:2011ws}.}.

In this case, we found an interesting similarity between the decay process of the black string \cite{Choptuik:2003qd, Lehner:2010pn, Lehner:2011wc} and the stochastic time evolution in the gauge theory.
In relation to it, we have three challenges.
First one is the investigation of the thermodynamical properties of the two dimensional super Yang-Mills theory (\ref{2dSYM}), which was partially done in Ref.~\cite{Catterall:2010fx}. 
Especially the decision of the order of the phase transition is important.
At the high temperature, the analysis of the one dimensional gauge theory  (\ref{BFSS}) may be valid, and there the $1/D$ expansion predicts that the phase transition at $T_{c1}$ is the second order transition followed by the GWW transition at $T_{c2}$ \cite{Mandal:2009vz}.
However, at the low temperature, the gravity calculation is valid and it predicts that the Gregory-Laflamme transition is first order \cite{Aharony:2004ig}.
Thus the order of the transition would depend on the parameters and it will be important to confirm it directly by investigating the model (\ref{2dSYM}).
In addition, the exploration of the relation between the cascade in the stochastic time evolution and the order of the phase transition may be also interesting.

The second challenge is the study of the real-time evolution of the uniform  configuration of $A_{x^1}$ in the two dimensional super Yang-Mills theory (\ref{2dSYM}) to compare the real-time evolution in the gravity.
Because of the appearance of the cascade decay in the thermodynamics of the one dimensional model (\ref{BFSS}), we expect that a similar decay pattern might appear in the real-time evolution, too.
One aim of this calculation is to see the fractal structure during the evolution, which was found in the black string decay \cite{Lehner:2010pn}.
This fractal structure is akin to a low viscosity fluid, and may indicate the appearance of the naked singularity.
Hence it plays a key role in  the breaking of the cosmic censorship hypothesis.
If we can observe the corresponding fractal structure in the gauge theory, it may be valuable in the context of the fluid/gravity correspondence \cite{Bhattacharyya:2008jc} and in the study of the resolution of the naked singularity through the quantum (finite-$N$) effects \cite{BMMW,  AlvarezGaume:2005fv, Liu:2004vy, AlvarezGaume:2006jg}.
(In the stochastic time evolution of the model (\ref{BFSS}), any clear fractal structure has not been observed. 
We need further numerical calculations at larger $N$ to conclude it.)

The last challenge is the numerical analysis of the decay process of the smeared black D0 brane solution (\ref{metric-D0}) in the IIA supergravity.
The authors in Refs.~\cite{Choptuik:2003qd, Lehner:2010pn, Lehner:2011wc} studied the decay of a 5 dimensional neutral black string in an asymptotically flat spacetime with $S^1$, which is distinct from the smeared D0 brane geometry.
To compare the real-time evolution in the gauge theory, it is important to study the D0 brane geometry.\\

{\em Acknowledgments---}
We would like to thank S.~Hashimoto, S.~Iso, G.~Mandal, J.~Nishimura, M.~Nozawa, Y.~Sakamura and S.~Tomizawa for useful discussions, and K.N.~Anagnostopoulos and S.~Nishida for technical supports. 
We would also like to thank G.~Mandal and S.~Minwalla for comments on the draft.
Numerical computations have been carried out on PC clusters at KEK, NTUA and YITP. 
 The work of T.A. is supported in part by Grant-in-Aid for Scientific
Research (No. 23740211) from JSPS.

\appendix

\section{General Multi-cut Solutions in the One-dimensional Gauge Theory}
\label{app-multi}

In section \ref{sec-Zm}, we derive $Z_m$ solution in the model (\ref{BFSS}).
In this appendix, we consider the derivation of general multi-cut solutions at high temperature $T \gg T_{c2}$.

The effective action (\ref{effective-2}) can be rewritten as
\begin{align} 
\frac{S_{\rm eff}(\alpha_i,\Delta)}{D}=&-\frac{\beta \Delta^4 N^2}{8 \tilde{\lambda}}
+\frac{1}{2} \sum_{i,j}^N \log \left[2\cosh(\beta \Delta)-2\cos(\alpha_i-\alpha_j  )  \right] 
\nonumber \\ &
-\frac{1}{2D} \sum_{i,j}^N \log \left( \sin^2 \left( \frac{\alpha_i-\alpha_j }{2} \right) \right) .
\label{effective}
\end{align} 
To derive an $m$-cut solution, 
we consider $m$ mobs of the eigenvalues, 
where the $l$-th mob consists of $N_l$ eigenvalues.
Here $\sum_{l=1}^m N_l =N$ and assume $N_l \sim O(N)$ and $m \ll N$. 
Then we decompose the eigenvalues in the $l$-th mob as
\begin{align} 
\alpha_i = \theta_l + \alpha^{(l)}_{j} , 
\quad {\rm for} \quad
\sum_{k=1}^{l-1} N_k < i  \le  \sum_{k=1}^{l} N_k,  
\qquad
j = i-   \sum_{k=1}^{l-1} N_k, \nonumber
\end{align} 
where $\theta_l$ and  $\alpha^{(l)}_{j}$ represent the centre of mass and the relative position of the $l$-th mob.  
$\alpha^{(l)}_{j}$ satisfies the constraint $\sum_{j=1}^{N_l}\alpha^{(l)}_{j}=0$, which is negligible at large $N$.
We assume $|\alpha_i^{(l)} | \ll |\theta_k-\theta_n| $ and $ |\alpha_i^{(l)} |\ll  \beta \Delta \ll 1$, and decompose the action (\ref{effective}) as 
\begin{align} 
&S_{\rm eff}(\alpha_i,\Delta)=S_{\theta}+S_{\Delta}+\sum_{l=1}^m S_{\alpha^{(l)}}, \quad \text{ with}
\label{action-decomp}
\\
&S_{\Delta}/DN^2 =-\frac{\beta \Delta^4 }{8 \tilde{\lambda}}
+ \sum_{l=1}^m \left(  \frac{N_l}{N}\right)^2 \log \left( \beta \Delta \right) + \cdots,
\nonumber \\
&S_{\theta}/D= \frac{1}{2} \sum_{k \ne l } N_k N_l \log \left(1- \cos \left( \theta_l-\theta_k\right) \right) + \cdots , 
\nonumber \\
&S_{\alpha^{(l)}} =\frac{N_lD}{ (\beta \Delta)^2}\sum_{i}^{N_l}  \left( \alpha^{(l)}_{i} \right)^2
-\frac{1}{2} \sum_{i,j}^{N_l}  
 \log \left( \alpha^{(l)}_{i}-\alpha^{(l)}_{j} \right)^2 + \cdots  
\nonumber .
\end{align}
Here $\alpha_i^{(l)}$ and $\theta_l$ appear only in $S_{\alpha^{(l)}}$ and $S_{\theta}$ respectively, and we can solve them independently.

First, we consider $S_{\alpha^{(l)}}$.
Then we see that $\alpha_i^{(l)}$ behave as a free Hermitian matrix model with a mass $M \equiv \sqrt{2D}/\beta \Delta$ \cite{Brezin:1977sv}.
Thus $\alpha_i^{(l)}$ compose a cut with a width $ 4 \beta \Delta/\sqrt{2D} $ and the assumption $|\alpha_i^{(l)} |\ll \beta \Delta$ is justified at large $D$.

Next we turn to $S_{\theta}$. 
$m$ equations of motion are derived from $S_{\theta}$,
\begin{align}
\sum_{k\neq l}^{m} \frac{N_k}{N} \cot\left( \frac{\theta_k-\theta_l}{2} \right) =0, \quad (l=1,\cdots,m) .
\label{eom-theta}
\end{align} 
Although these equations are complicated, we would obtain a solution $\{\theta_l \}$, which satisfies $\theta_k \neq \theta_l$, for a given set of $N_l$ at least numerically\footnote{As far as we calculated, we always found at least one numerical solution of eq.(\ref{eom-theta}) for a given set of $N_l$.}.

Lastly we consider $S_{\Delta}$. (Additional potentials for $\Delta$ arise by integrating $\alpha_i^{(l)}$ in $S_{\alpha^{(l)}}$ but they are negligible at large $D$.)
$S_{\Delta}$ has a unique saddle point
\begin{align} 
\Delta = \left(\frac{2\tilde{\lambda}}{ \beta} \sum_{l=1}^m \left(\frac{N_l}{N} \right)^2 \right)^{1/4}.
\end{align} 
This result implies that the assumptions $\beta \Delta \ll1 $ and
$|\alpha_i^{(l)} | \ll |\theta_k-\theta_n| $ are valid at sufficiently high temperature. 
In this way, we obtain the multi-cut solutions in the model (\ref{BFSS}). 

Lastly we evaluate the observables in this solution.
The value of the effective action is given by 
\begin{align} 
\frac{S_m}{\beta DN^2}=&\frac{S_{\Delta}}{\beta DN^2}+O(1/D)
=   \sum_{l=1}^m \left(  \frac{N_l}{N}\right)^2  \frac{ \log \left( \beta \Delta \right)}{\beta} +\cdots
\label{energy}
\end{align} 
at the high temperature.
The eigenvalue density becomes
\begin{align} 
\rho(\alpha)=&\sum_{l=1}^m \frac{N_l}{N} \rho_l(\alpha-\theta_l), 
\label{rho-m}
\\ 
\rho_l (\alpha)=& \frac{1}{N_l} \sum_{i=1}^{N_l} \delta(\alpha-\alpha_i^{(l)}) =\frac{M^2}{2\pi} {\rm Re}\left( \sqrt{\frac{4}{M^2}-\alpha^2} \right)
 \nonumber.
\end{align} 
One interesting feature of this density is that the widths of each cut are commonly given by $4/M$.
Indeed, in the numerical calculation of the stochastic evolution in Fig.~\ref{fig-cascade}, we observe that the widths of the cuts of steady states seem common.

Note that $\Delta$ and $S_m$ are determined by $N_l$ and $\beta$, and do not depend on $\theta_l$.
It may explain the reason that, in the stochastic evolution in Fig.~\ref{fig-ZmMC}, $\Delta$ remains almost constant until around $s=2200$, whereas $\{u_n \}$ start varying from around $s=1000$.
In this evolution, the two mobs gradually approach each other, and finally they merge into one mob.
Since $\{u_n \}$ are sensitive to the angle of these two mobs, they are varying during the evolution.
On the other hand, since the numbers of the eigenvalues in each mob do not change until they merge, $\Delta$ may remain constant.

\section{Details of the Monte Carlo Calculation}
\label{app-MC}

We explain the details of the stochastic time evolution through the Monte Carlo calculation of the model (\ref{BFSS}), which we employ in Sec.~\ref{sec-cascade}.

\paragraph{Action:}
We adopt the same action introduced in Appendix B of Ref.~\cite{Kawahara:2007fn}. 

\paragraph{Random number $r$:}
$r$ in ``1.'' in Sec.~\ref{subsec-stochastic} is taken $r=\tau N(0,1)$, where $\tau$ is a small number and $N(0,1)$ is a random
number with standard normal distribution. Typically, we take $\tau =
0.01$.

\paragraph{Adjoint scalar:}
In order to achieve the equilibrium in the adjoint scalar $Y^I$ at each step  as argued ``4.'' in Sec.~\ref{subsec-stochastic}, we repeat the heatbath
updating of the scalar fields sufficiently many times.
Here we observe that, if the repetition number is not enough, the stochastic evolution does depend on the lattice spacing.
Smaller lattice spacing requires more repetitions to achieve the equilibrium.
Typically we take the lattice spacing 0.05 and repeat 5 times at each step.
About this lattice spacing, see Ref.~\cite{Kawahara:2007nw}.
We have confirmed that this repetition number is sufficient at this lattice spacing.


\begin{thebibliography}{999}

\bibitem{Gross:1980br}
  D.~J.~Gross, R.~D.~Pisarski and L.~G.~Yaffe,
  ``QCD and Instantons at Finite Temperature,''
  Rev.\ Mod.\ Phys.\  {\bf 53} (1981) 43.
 



\bibitem{Creutz:1984mg}
  M.~Creutz,
  ``Quarks, Gluons And Lattices,''
  Cambridge, Uk: Univ. Pr. ( 1983) 169 P. ( Cambridge Monographs On Mathematical Physics)


\bibitem{Maldacena:1997re}
  J.~M.~Maldacena,
  ``The Large N limit of superconformal field theories and supergravity,''
  Adv.\ Theor.\ Math.\ Phys.\  {\bf 2 } (1998)  231-252.
  [hep-th/9711200].

\bibitem{Witten:1998zw}
  E.~Witten,
  ``Anti-de Sitter space, thermal phase transition, and confinement in gauge theories,''
  Adv.\ Theor.\ Math.\ Phys.\  {\bf 2} (1998) 505
  [hep-th/9803131].

\bibitem{Gross:1998gk}
  D.~J.~Gross and H.~Ooguri,
  ``Aspects of large N gauge theory dynamics as seen by string theory,''
  Phys.\ Rev.\  D {\bf 58}, 106002 (1998)
  [arXiv:hep-th/9805129].

\bibitem{Aharony:1998qu}
  O.~Aharony and E.~Witten,
  ``Anti-de Sitter space and the center of the gauge group,''
  JHEP {\bf 9811} (1998) 018
  [hep-th/9807205].

\bibitem{Mandal:2011ws}
  G.~Mandal and T.~Morita,
  ``Gregory-Laflamme as the confinement/deconfinement transition in holographic QCD,''
  JHEP {\bf 1109} (2011) 073
  [arXiv:1107.4048 [hep-th]].

\bibitem{Aharony:2005bm} 
  O.~Aharony, S.~Minwalla and T.~Wiseman,
  ``Plasma-balls in large N gauge theories and localized black holes,''
  Class.\ Quant.\ Grav.\  {\bf 23}, 2171 (2006)
  [hep-th/0507219].


\bibitem{Luscher:1982ma}
  M.~Luscher,
  ``Some Analytic Results Concerning The Mass Spectrum Of Yang-Mills Gauge
  Nucl.\ Phys.\  B {\bf 219} (1983) 233.


\bibitem{Mandal:2011hb}
  G.~Mandal and T.~Morita,
  ``Phases of a two dimensional large N gauge theory on a torus,''
  Phys.\ Rev.\ D {\bf 84} (2011) 085007
  [arXiv:1103.1558 [hep-th]].


\bibitem{'tHooft:1973jz}
  G.~'t Hooft,
  ``A Planar Diagram Theory for Strong Interactions,''
  Nucl.\ Phys.\ B {\bf 72} (1974) 461.




\bibitem{Aharony:2004ig}
  O.~Aharony, J.~Marsano, S.~Minwalla and T.~Wiseman,
  ``Black hole-black string phase transitions in thermal 1+1-dimensional supersymmetric Yang-Mills theory on a circle,''
  Class.\ Quant.\ Grav.\  {\bf 21}, 5169 (2004)
  [arXiv:hep-th/0406210].


\bibitem{Aharony:2005ew} 
  O.~Aharony, J.~Marsano, S.~Minwalla, K.~Papadodimas, M.~Van Raamsdonk and T.~Wiseman,
  ``The Phase structure of low dimensional large N gauge theories on Tori,''
  JHEP {\bf 0601}, 140 (2006)
  [hep-th/0508077].
  
 
\bibitem{Mandal:2009vz}
  G.~Mandal, M.~Mahato, T.~Morita,
  ``Phases of one dimensional large N gauge theory in a 1/D expansion,''
  JHEP {\bf 1002}, 034 (2010).
  [arXiv:0910.4526 [hep-th]].

\bibitem{Itzhaki:1998dd}
  N.~Itzhaki, J.~M.~Maldacena, J.~Sonnenschein and S.~Yankielowicz,
  ``Supergravity and the large N limit of theories with sixteen
  supercharges,''
  Phys.\ Rev.\  D {\bf 58} (1998) 046004
  [arXiv:hep-th/9802042].


\bibitem{Harmark:2004ws}
  T.~Harmark and N.~A.~Obers,
  ``New phases of near-extremal branes on a circle,''
  JHEP {\bf 0409} (2004) 022
  [hep-th/0407094].


\bibitem{Hotta:1998en}
  T.~Hotta, J.~Nishimura and A.~Tsuchiya,
  ``Dynamical aspects of large N reduced models,''
  Nucl.\ Phys.\  B {\bf 545} (1999) 543
  [arXiv:hep-th/9811220].


\bibitem{Choptuik:2003qd}
  M.~W.~Choptuik, L.~Lehner, I.~Olabarrieta, R.~Petryk, F.~Pretorius and H.~Villegas,
  ``Towards the final fate of an unstable black string,''
  Phys.\ Rev.\ D {\bf 68} (2003) 044001
  [gr-qc/0304085].
 
\bibitem{Lehner:2010pn}
  L.~Lehner and F.~Pretorius,
  ``Black Strings, Low Viscosity Fluids, and Violation of Cosmic Censorship,''
  Phys.\ Rev.\ Lett.\  {\bf 105} (2010) 101102
    [arXiv:1006.5960 [hep-th]].


\bibitem{Lehner:2011wc}
  L.~Lehner and F.~Pretorius,
  ``Final State of Gregory-Laflamme Instability,''
  arXiv:1106.5184 [gr-qc].

\bibitem{Gregory:1994bj}
  R.~Gregory and R.~Laflamme,
  ``The Instability of charged black strings and p-branes,''
  Nucl.\ Phys.\  B {\bf 428} (1994) 399
  [arXiv:hep-th/9404071].

\bibitem{BMMW}
P.~Basu, G.~Mandal, T.~Morita and S.~Wadia,
work in progress


\bibitem{Manton:1979kb}
  N.~S.~Manton,
  ``A New Six-Dimensional Approach to the Weinberg-Salam Model,''
  Nucl.\ Phys.\ B {\bf 158} (1979) 141.
  
\bibitem{Fairlie:1979at}
  D.~B.~Fairlie,
  ``Higgs' Fields and the Determination of the Weinberg Angle,''
  Phys.\ Lett.\ B {\bf 82} (1979) 97.
  

\bibitem{Hosotani:1983xw}
  Y.~Hosotani,
  ``Dynamical Mass Generation by Compact Extra Dimensions,''
  Phys.\ Lett.\ B {\bf 126} (1983) 309.

\bibitem{Unsal:2010qh}
  M.~Unsal and L.~G.~Yaffe,
  ``Large-N volume independence in conformal and confining gauge theories,''
  JHEP {\bf 1008} (2010) 030
  [arXiv:1006.2101 [hep-th]].

\bibitem{Brezin:1977sv}
  E.~Brezin, C.~Itzykson, G.~Parisi and J.~B.~Zuber,
  ``Planar Diagrams,''
  Commun.\ Math.\ Phys.\  {\bf 59} (1978) 35.


  
\bibitem{Gross:1980he} 
  D.~J.~Gross and E.~Witten,
  ``Possible Third Order Phase Transition in the Large N Lattice Gauge Theory,''
  Phys.\ Rev.\ D {\bf 21}, 446 (1980).

\bibitem{Wadia:1980cp}
  S.~R.~Wadia,
  ``N = infinity PHASE TRANSITION IN A CLASS OF EXACTLY SOLUBLE MODEL LATTICE GAUGE THEORIES,''
  Phys.\ Lett.\ B {\bf 93} (1980) 403.

\bibitem{Jurkiewicz:1982iz}
  J.~Jurkiewicz and K.~Zalewski,
  ``Vacuum Structure Of The U(n $\to$ Infinity) Gauge Theory On A Two-dimensional Lattice For A Broad Class Of Variant Actions,''
  Nucl.\ Phys.\ B {\bf 220} (1983) 167.
    
\bibitem{Mandal:1989ry}
  G.~Mandal,
  ``Phase Structure Of Unitary Matrix Models,''
  Mod.\ Phys.\ Lett.\ A {\bf 5} (1990) 1147.


\bibitem{Aharony:2003sx} 
  O.~Aharony, J.~Marsano, S.~Minwalla, K.~Papadodimas and M.~Van Raamsdonk,
  ``The Hagedorn - deconfinement phase transition in weakly coupled large N gauge theories,''
  Adv.\ Theor.\ Math.\ Phys.\  {\bf 8}, 603 (2004)
  [hep-th/0310285].
  
\bibitem{AlvarezGaume:2005fv}
  L.~Alvarez-Gaume, C.~Gomez, H.~Liu and S.~Wadia,
  ``Finite temperature effective action, AdS(5) black holes, and 1/N expansion,''
  Phys.\ Rev.\ D {\bf 71} (2005) 124023
  [hep-th/0502227].
  
    
\bibitem{Liu:2004vy}
  H.~Liu,
  ``Fine structure of Hagedorn transitions,''
  hep-th/0408001.
  
  

\bibitem{AlvarezGaume:2006jg}
  L.~Alvarez-Gaume, P.~Basu, M.~Marino and S.~R.~Wadia,
  ``Blackhole/String Transition for the Small Schwarzschild Blackhole of AdS(5)x S**5 and Critical Unitary Matrix Models,''
  Eur.\ Phys.\ J.\ C {\bf 48} (2006) 647
  [hep-th/0605041].

\bibitem{Kawahara:2007fn}
  N.~Kawahara, J.~Nishimura and S.~Takeuchi,
  ``Phase structure of matrix quantum mechanics at finite temperature,''
  JHEP {\bf 0710} (2007) 097
  [arXiv:0706.3517 [hep-th]].

\bibitem{Barbon:1998cr}
  J.~L.~F.~Barbon, I.~I.~Kogan and E.~Rabinovici,
  ``On stringy thresholds in SYM / AdS thermodynamics,''
  Nucl.\ Phys.\ B {\bf 544} (1999) 104
  [hep-th/9809033].



\bibitem{Hollands:2012sf}
  S.~Hollands and R.~M.~Wald,
  arXiv:1201.0463 [gr-qc].
  
  
\bibitem{Catterall:2010fx}
  S.~Catterall, A.~Joseph and T.~Wiseman,
  ``Thermal phases of D1-branes on a circle from lattice super Yang-Mills,''
  JHEP {\bf 1012} (2010) 022
  [arXiv:1008.4964 [hep-th]].
  
  
  
\bibitem{Bhattacharyya:2008jc}
  S.~Bhattacharyya, V.~Hubeny, S.~Minwalla and M.~Rangamani,
  ``Nonlinear Fluid Dynamics from Gravity,''
  JHEP {\bf 0802} (2008) 045
  [arXiv:0712.2456 [hep-th]].
  


\bibitem{Kawahara:2007nw} 
  N.~Kawahara, J.~Nishimura and S.~Takeuchi,
  ``Exact fuzzy sphere thermodynamics in matrix quantum mechanics,''
  JHEP {\bf 0705}, 091 (2007)
  [arXiv:0704.3183 [hep-th]].
  
  
  

\end{thebibliography}
\end{document}